%% file: paper.tex
\documentclass[usenatbib]{mn2e}
\usepackage{amsmath}
\usepackage{natbib}
\usepackage{epsfig}
\usepackage{txfonts}
\usepackage{epstopdf}

\input{Befehle}

\newcommand{\matrixf}[1]{\boldsymbol{\mathsf{#1}}}

\newcommand{\beamav}[1]{\left< #1\right>}
\newcommand{\TeSZ}{T_{\rm e, SZ}}
\newcommand{\betacparaSZ}{\beta_{\rm c, \parallel, SZ}}
\newcommand{\betacperpSZ}{\beta_{\rm c, \perp, SZ}}
\newcommand{\betacpara}{\beta_{\rm c, \parallel}}

\newcommand{\MJysr}{{\rm MJy\,sr^{-1}}}
\newcommand{\keV}{{\rm keV}}
\newcommand{\Teh}{\hat{T}_{\rm e}}
\newcommand{\Mpc}{{\rm Mpc}}
\newcommand{\kpc}{{\rm kpc}}
\newcommand{\GHz}{{\rm GHz}}

\usepackage{color}
\newcommand{\changeJ}[1]{{#1}}
\newcommand{\changeC}[1]{{#1}}
\usepackage{hyperref}
\usepackage{grffile}
\usepackage{graphics}
\usepackage{booktabs}

\title[SZ signal processing]
{Sunyaev-Zeldovich signal processing and temperature-velocity moment 
method for individual clusters}

\author[Chluba et al.]{
  Jens Chluba$^{1,2}$\thanks{E-mail: jchluba@pha.jhu.edu}, 
  Eric Switzer$^{1, 3}$, 
  Kaylea Nelson$^{4}$, 
  Daisuke Nagai$^{4, 5, 6}$
  \\ \\
$^{1}$ Canadian Institute for Theoretical Astrophysics, 60 St. George Street,
Toronto, ON M5S 3H8, Canada
\\
$^{2}$ Johns Hopkins University, Department of Physics and Astronomy, Bloomberg 439, 
3400 N. Charles St., Baltimore, MD 21218
\\
$^{3}$ Kavli Institute for Cosmological Physics, University of Chicago, 
933 East 56th Street, Chicago, IL 60637, USA
\\
$^{4}$ Department of Astronomy, Yale University, New Haven, CT 06520, U.S.A.
\\
$^{5}$ Department of Physics, Yale University, New Haven, CT 06520, U.S.A.
\\
$^{6}$ Yale Center for Astronomy \& Astrophysics, Yale University, New Haven, CT 06520, U.S.A.
}

\begin{document}

\date{Accepted 2013 January 16. Received 2013 January 09; in original form 2012 November 14}

\maketitle

\begin{abstract}
Future high resolution, high sensitivity Sunyaev-Zeldovich (SZ) observations of individual clusters will provide an exciting opportunity to \changeJ{answer} specific questions about the dynamical state of the intra-cluster medium (ICM). 
In this paper we develop a new method that clearly shows the connection of the SZ signal with the \changeJ{underlying} cluster model. We include relativistic temperature and kinematic corrections in the single-scattering approximation, allowing studies of hot clusters.  In our approach, particular {\it moments} of the temperature and velocity field along the line-of-sight determine the precise spectral shape and morphology of the SZ signal. 
We illustrate how to apply our method to different cluster models, highlighting parameter degeneracies and instrumental effects that are important for interpreting future 
high-resolution SZ data. 
Our analysis shows that line-of-sight temperature variations can introduce significant biases in the derived SZ temperature and peculiar velocity. We furthermore discuss how the position of the SZ null is affected by the cluster's temperature and velocity structure. Our computations indicate that the SZ signal around the null alone is rather insensitive to different cluster models and that high frequency channels add a large leverage in this respect.
We also apply our method to recent high sensitivity SZ data of the Bullet cluster, showing how the results can be linked to line-of-sight variations in the electron temperature.
The tools developed here as part of {\sc SZpack} should be useful for analyzing high-resolution SZ data and computing SZ maps from simulated clusters.
\end{abstract}
%------------------------------------------------------------


\section{Introduction}
\label{sec:Intro}
%------------------------------------

Since the early measurements of the SZ effect in the 80's and 90's \citep[e.g.,][]{Birkinshaw1984, Birkinshaw1991, Lamarre1998, Hughes1998} the observational possibilities in the microwave band have evolved at an impressive rate. Today the thermal SZ (thSZ) effect \citep{Zeldovich1969} is routinely detected for several hundred galaxy clusters out to redshift $z\simeq 1$ \citep{Benson2004, Marriage2011, Williamson2011, AdeESZCS}, and in some cases evidence for the presence of the smaller kinematic SZ (kSZ) effect \citep{Sunyaev1980} is found \citep{Benson2003, Korngut2011, Mroczkowski2012, Hand2012}.

SZ cluster observations are a powerful cosmological tool \changeJ{\citep[see][for overview]{Rephaeli1995ARAA, Birkinshaw1999, Carlstrom2002, Komatsu2002}}, but to realize their full constraining power for precision cosmology it is critical to understand the structure and evolution of the intra-cluster medium (ICM) in more detail.
Already now a number of high-resolution SZ experiments, including ALMA\footnote{Atacama Large Millimeter/submillimeter Array}, CARMA\footnote{Combined Array for Research in Millimeter-wave Astronomy}, CCAT\footnote{Cornell Caltech Atacama Telescope}, and MUSTANG\footnote{MUltiplexed Squid TES Array at Ninety GHz}, are underway or planned, promising a dramatic increase in sensitivities, spatial resolution, and spectral coverage over the next few years.
Current generation high-resolution SZ observations of individual clusters have already revealed rich phenomena (e.g., shocks, substructures, relativistic particles) in the atmospheres of merging clusters \changeJ{\citep[e.g.,][]{Komatsu2001, Kitayama2004, Colafrancesco2011, Korngut2011, ElGordo2012, Mroczkowski2012, Zemcov2012, Prokhorov2012}}. 
Upcoming SZ experiments should enable a host of exciting measurements of important physical processes that shape the properties of the ICM and their evolution, including the electron temperature \citep{Pointecouteau1998, Hansen2002}, the peculiar velocity of the cluster, internal bulk and turbulent gas motions \citep[e.g.,][]{Chluba2002, Nagai2003, Sunyaev2003, Diego2003}, and non-equilibrium electrons produced by merger and accretion shocks \citep{Markevitch2007, Rudd2009}.

These encouraging prospects also raise a number of important problems that must be addressed before the rich information contained in the future SZ data can be fully exploited.
One is simply related to the precise and fast computation of the SZ signal given basic parameters of the scattering medium, such as the Thomson scattering optical depth, $\tau$, the electron temperature, $\Te$, and bulk velocity, $\betac$, while accounting for relativistic temperature and kinematic corrections. 
\changeJ{Previously, this} issue has been addressed by several groups \citep{Rephaeli1995, Challinor1998, Itoh98, Sazonov1998, Nozawa1998SZ, Challinor1999, Chluba2005b} by means of Taylor expansions for the SZ signal. While the evaluation of these expansions is very fast, they are limited to rather low temperature gas (Fig.~\ref{fig:precision_Itoh_CNSN} shows that for $\Te\gtrsim 13\,\keV$ this approach breaks down).
One alternative is direct numerical integration of the Boltzmann collision term making use of the symmetries of the scattering problem \citep{Wright1979, Dolgov2001, Nozawa2009, Poutanen2010}, but this is time-consuming and not well-suited for extensive parameter estimations or computations of the SZ signal from simulated clusters.
On the other hand, a fast but not as precise and flexible approach is simple tabulation of the SZ signal or the use of convenient fitting function \citep{Nozawa2000fitting, Itoh2004fittingII, Shimon2004}. 

Recently, \citet[][CNSN in the following]{ChlubaSZpack} developed a method in the middle of these extremes. In their work, a new set of frequency-dependent basis functions was computed numerically to allow very fast and precise calculation of the SZ signal. The basis functions are informed by the underlying physics of the scattering problem and thus are ideally suited for future SZ signal processing and parameter estimation. 
The associated routines are part of {\sc SZpack}\footnote{\url{www.Chluba.de/SZpack}; (now also including Python bindings)}. 
However, several extensions are required. Firstly, so far high precision (relative accuracy $\simeq 0.001\%$) calculations with {\sc SZpack} were limited to $\Te \lesssim 25\, \keV$. 
This problem can be easily overcome using the method of CNSN by appropriate extensions of the basis functions, as we explain in Sect.~\ref{sec:numberofmoments}.
With {\sc SZpack v1.1}, which is presented here, it is now possible to compute the SZ signal for $\Te \lesssim 75\,\keV$ and $\betac \lesssim 0.01$ to $\simeq 0.001\%$ relative precision at practically no computational cost. 
This precision and range of parameters covers all physically relevant cases and hence provides an important preparation for SZ parameter estimation without significant limitations.

Secondly, line-of-sight variations of the temperature and velocity field (with {\it any} of the aforementioned methods) can only be accounted for by means of additional 1-dimensional integrals; this again makes extensive SZ parameter estimation expensive.
Especially when computing the SZ signal from cluster simulations the problem becomes very demanding, even if evaluation for single gas parameters ($\tau$, $\Te$, and $\betac$) is extremely fast.
Here we reformulate the representation of the SZ signal to overcome this limitation. 
We utilize that in the single-scattering approximation, frequency-dependent terms can be separated from temperature- and velocity-dependent contributions (Sect.~\ref{sec:Moments}).
This implies that the SZ signal for a given cluster model can be calculated using appropriate {\it moments} of the temperature and velocity field.
While this means that a finite number of 1-dimensional integrals along different lines-of-sight have to be evaluated, this separation still greatly reduces the computational burden because afterward the SZ signal at any frequency can be computed as simple matrix multiplication.

While the method developed in Sect.~\ref{sec:Moments} is both precise and fast, delivering quasi-exact results for the SZ signal through different lines-of-sight for {\it any} cluster atmosphere, in the future analysis of high resolution, high sensitivity SZ data another simplification is possible. In Sect.~\ref{sec:SZclusters}, we show that for typical cluster models the smoothness of the temperature and velocity profiles allows minimizing the number of parameters needed to accurately describe the SZ signal, resulting in a second set of moments that are related to the line-of-sight temperature and velocity dispersions and higher order statistics (see Eq.~\eqref{sec:Signal_eq_tot_series_velocity} for instance). The associated expansion of the SZ signal around the mean becomes perturbative and the number of moments needed to describe the SZ signal depends directly on the observational sensitivity.

Our formulation furthermore allows direct separation of frequency-dependent from spatially varying terms, providing a clear link between {\it morphological changes} of the SZ signal and cluster parameters. 
For example, the presence of large-scale, post-merger cluster rotation can introduce a bipolar kSZ signal, which is related to a spatially varying average line-of-sight velocity. The superposition of thSZ with this rotational kSZ implies small frequency-dependent changes of the clusters morphology \citep{Chluba2002}. Similarly, variations of the electron temperature along the line-of-sight introduce morphological effects \citep[as also pointed out more recently by][]{Prokhorov2011}, and as we explain here, spatial variations of temperature and velocity moments are the source of these morphological changes. 
The moments therefore constitute the main observables of high-resolution, high-sensitivity SZ observations and their interpretation is the main challenge for future SZ parameter estimation and in the reconstruction of the cluster's temperature and velocity structure.

Armed with these tools, we address a number of questions that are related to the effect of temperature and velocity variations on the SZ signal, with particular focus on parameter degeneracies and instrumental aspects. For example, we explicitly discuss the effect of angular resolution and frequency filters on the SZ signal, as well as different corrections to the location of the SZ null.
All these aspects, if ignored, lead to biases in the deduced cluster parameters. 
We illustrate this for several examples, using both mock SZ measurements as well as recent SZ data.
We furthermore develop several tools for SZ parameter estimation which are now part of {\sc SZpack}.
These should be useful for computing the SZ signal from cluster simulations and in the analysis of future high resolution, high sensitivity SZ measurements.

\section{Computing the SZ signal using temperature and velocity moments}
\label{sec:Moments}
%------------------------------------
In this section we introduce the new temperature-velocity moment method to compute the SZ signal for general cluster atmospheres. This section is rather technical and mainly for readers interested in the computational details.

The SZ effect is caused by the scattering of CMB photons by moving electrons. For a small volume element of scattering electrons the SZ signal only depends on the electron temperature, $\Te$, their total bulk velocity, $\vbetac$, the direction cosine of this velocity with respect to the line-of-sight\footnote{In the following bold font denotes 3-dimensional vectors and an additional hat means that it is normalized to unity.}, $\muc=\vbetach\cdot\vgh$, and the Thomson optical depth of the scattering volume element, $\Delta \tau$.
Both $\Te$ and $\Delta \tau$ are defined in the rest frame of the scattering volume element, while $\vbetac$ is defined with respect to the CMB rest frame.

For now, we shall assume that the observer is at rest in the CMB frame.
The conversion of the aforementioned parameters to the SZ signal can be expressed as $\Delta I(x)=\Delta \tau \,F_{\rm SZ}(x, \Te, \betac, \muc)$, where $F_{\rm SZ}$ is a non-linear function and $x=h\nu/kT_0$ with CMB temperature $T_0=2.726\,{\rm K}$ \citep{Fixsen1996, Fixsen2002}.
\changeC{The change in the CMB intensity} can be further rewritten as 
%--------------
\beal
\label{eq:SZ_signal}
\Delta I(x)&\approx \Delta \tau \,\changeC{I_{\rm o}\,x^3} \left[Y(x, \Te) + \betacsq M(x, \Te) 
\right.
\nonumber\\
&\quad\qquad
\left.
+ \betac P_1(\muc) \,D(x, \Te) 
+ \betacsq P_2(\muc) \,Q(x, \Te) \right],
\end{align}
%--------------
where \changeC{$I_{\rm o}=(2h/c^2) (kT_0/h)^3\approx 270\,\MJysr$ and} $P_l(x)$ denote Legendre polynomials. The term $Y(x, \Te)$ describes the purely thermal SZ effect with temperature corrections included, while terms $\propto \betac$ are related to kinematic effects.

From previous analysis of the SZ effect, it is furthermore clear that the functions $Y(x, \Te)$, $M(x, \Te)$, $D(x, \Te)$ and $Q(x, \Te)$ can all be described using an appropriate set of frequency-dependent basis functions and temperature-dependent coefficients, where the latter encode the spatial dependence. 
The integrated SZ signal along a given line-of-sight is therefore determined by appropriate moments of the cluster's temperature and velocity field. These contain the desired information about the cluster gas and structure; as such they define the observables of the SZ measurement, and the aim will be to use the moments to learn about the cluster gas.
%%
%\changeJ{Below we give a description for the functions, $Y(x, \Te)$, $M(x, \Te)$, $D(x, \Te)$ and $Q(x, \Te)$. At low temperatures, an asymptotic expansion in terms of $k\Te\ll \me c^2$ is applicable, while at higher temperature the basis of CNSN is used. The former breaks down at high temperatures, while the latter becomes non-perturbative at low temperatures (see Fig.~\ref{fig:precision_Itoh_CNSN}).}

\subsection{Low temperature gas ($k\Te\lesssim 10\,{\rm keV}$)}
As discussed in \citet{ChlubaSZpack}, for electron gas temperatures $k\Te\lesssim 5\,{\rm keV}-10\,{\rm keV}$ an asymptotic expansion of the Boltzmann collision term \citep[similar to][]{Challinor1998, Itoh98, Sazonov1998, Nozawa1998SZ, Challinor1999} can be used to represent the SZ signal with high precision.
From Eq. (25) of CNSN, up to some specified correction order, $k_{\rm max}$, in the electron temperature, it therefore follows (see Appendix~\ref{app:der_asym} for more details):
%--------------
\bsub
\label{eq:SZ_signal_functions}
\beal
Y^{\rm low}(x, \Te) &= \sum_{k=0}^{k_{\rm max}} Y_k(x) \, \The^{k+1}
\\
M^{\rm low}(x, \Te) &= \frac{1}{3}\mathcal{M}(x)
+
\sum_{k=0}^{k_{\rm max}} M^{\rm low}_k(x) \,\The^{k+1}
\\
D^{\rm low}(x, \Te)&=\mathcal{G}(x) +  \sum_{k=0}^{k_{\rm max}}  
D^{\rm low}_k(x) \,\The^{k+1}
\\
Q^{\rm low}(x, \Te)&=\frac{11}{30}\,\mathcal{Q}(x)
+
\sum_{k=0}^{k_{\rm max}} Q^{\rm low}_k(x)\, \The^{k+1},
\end{align}
\esub
%--------------
with $\The=k\Te/\me c^2$ , $\mathcal{M}(x)=Y_0(x)+\mathcal{G}(x)$, $\mathcal{G}(x)=xe^x/[e^x-1]^2$, and $\mathcal{Q}=x \,\mathcal{G}\coth(x/2)$.
The functions $Y_k$ are defined as in CNSN, while $M^{\rm low}_k$, $D^{\rm low}_k$, and $Q^{\rm low}_k$ are given by Eq.~\eqref{eq:SZ_signal_functions_defs_explicit}.
To give an example, $Y_0=\mathcal{Q}-4\mathcal{G}$ describes the usual (non-relativistic) thSZ effect \citep{Zeldovich1969}, while the term $\propto \mathcal{G}$ is related to the kSZ effect \citep{Sunyaev1980}.
Both $\mathcal{M}(x)$ and $\mathcal{Q}(x)$ describe the lowest order kinematic terms $\propto \betacsq$.

For a fixed line-of-sight through the cluster medium, the total SZ signal is determined by integration over $\Delta \tau$. With the decomposition given above it is convenient to introduce the following line-of-sight temperature and velocity moments:
%--------------
\beal
\label{eq:SZ_signal_y_parameters}
y^{(k)}&= \int \The^{k+1} \id \tau,
&
b^{(k)}_0&= \int \betacsq \The^{k} \id \tau,
\\\nonumber
b^{(k)}_1&= \int \betac P_1(\muc) \The^{k} \id \tau,
&b^{(k)}_2&= \int \betacsq P_2(\muc) \The^{k} \id \tau,
\end{align}
%--------------
where the integrals are carried out in the cluster frame, with the condition $\Te\leq T_{\rm e, low}$ for $k>0$ and for $y^{(0)}$. 
Here $y^{(k)}$ denotes the generalized $y$-parameters, while $b^{(k)}_i$ take into account the effect of the clusters global and internal gas motion. For example, $y^{(0)}=\int \The \id \tau$ is the usual line-of-sight $y$-parameter or average thermal pressure of the electrons, while $b^{(0)}_1$ (no temperature dependence) is proportional to the average velocity of the cluster medium along the line-of-sight weighted by the electron number density.
The optimal values for $T_{\rm e, low}$ and $k_{\rm max}$ depend on the required precision and will be specified below (see Sect.~\ref{sec:numberofmoments}).
With Eq.~\eqref{eq:SZ_signal_y_parameters} we can now define the moment vector
%--------------
\beal
\vek{m}^T_{\rm low}&=(\vek{y}^T, \vek{b}_1^T, \vek{b}_0^T, \vek{b}_2^T)
\nonumber\\
\vek{y}^T&=(y^{(0)}, ..., y^{(k_{\rm max})})
\\\nonumber
\vek{b}_i^T&=(b^{(0)}_i, ..., b_i^{(k_{\rm max}+1)}),
\end{align}
%--------------
where we arranged the entries of $\vek{m}_{\rm low}$ with respect to orders in $\betac$. Notice that the dimensions of the vectors $\vek{b}_i$ in principle can differ from $k_{\rm max}+1$.
In particular, for the velocity moments usually fewer terms in the electron temperature are required to describe the SZ signal accurately, since they only lead to very small corrections \citep[e.g., see][]{ChlubaSZpack}.

If we assume that the SZ signal is observed at $m$ frequencies, $\{x_i\}$, we can furthermore introduce the signal vector for the contribution of the low temperature gas, $\vek{S}^T_{\rm low}=(\Delta I_{\rm low}(x_1), ..., \Delta I_{\rm low}(x_m))$. This defines the matrix equation
%--------------
\beal
\label{sec:Signal_eq}
\vek{S}_{\rm low}&=\matrixf{F}_{\rm low} \vek{m}_{\rm low},
\end{align}
%--------------
where each row of the matrix $\matrixf{F}_{\rm low}$ reads 
%--------------
\beal
(\matrixf{F}_{\rm low})_{i}=\changeC{I_{\rm o}\,x_i^3}\left(\vek{Y}^T_{\rm low}, 
\mathcal{G}, \vek{D}^T_{\rm low}, 
\frac{1}{3} \mathcal{M}, \vek{M}^T_{\rm low}, 
\frac{11}{30}\mathcal{Q}, \vek{Q}^T_{\rm low}\right),
\end{align}
%--------------
with $\vek{Y}^T_{\rm low}=(Y_0, ..., Y_{k_{\rm max}})$, 
$\vek{D}^T_{\rm low}= (D^{\rm low}_0, ..., D^{\rm low}_{k_{\rm max}})$, and so on.
For the $i^{\rm th}$ row of $\matrixf{F}_{\rm low}$ the frequency-dependent functions evidently have to be evaluated at the required $x_i$.
The SZ signal caused by low temperature gas can therefore be computed as a simple matrix operation once the low temperature-velocity moment vector, $\vek{m}_{\rm low}$, is known. The columns of the moment matrix, $\matrixf{F}_{\rm low}$, are the values of the basis functions at the required frequencies.

\subsection{High temperature gas ($k\Te\gtrsim 10\,{\rm keV}$)}
\begin{figure}
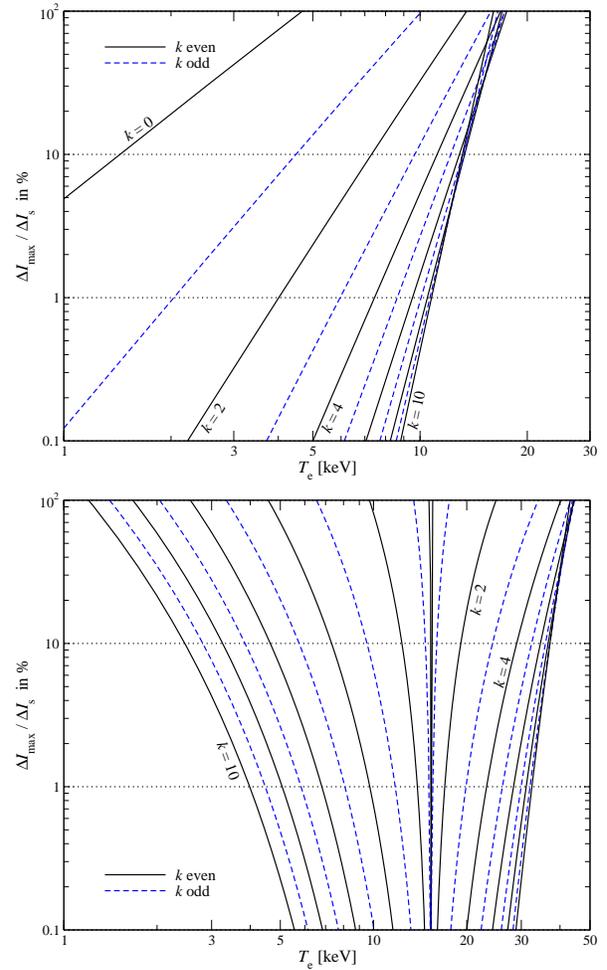

\centering
\includegraphics[width=0.92\columnwidth]{./eps/rec.moments.eps}
\\[1mm]
\includegraphics[width=0.92\columnwidth]{./eps/rec.moments.CNSN.eps}
\caption{Range of convergence for different approximations. The upper panel shows the results obtained with the asymptotic expansion, while for the lower panel the basis CNSN with $\theta_{\rm e, 0}=0.03$ was used. The maximal departure for $0.1\leq x\leq 30$ is expressed in units of $\Delta I_{\rm s}\simeq 0.013\, {\rm MJy\,sr^{-1}}$ and $\tau=0.01$ was assumed for all cases.}
\label{fig:precision_Itoh_CNSN}
\end{figure}

For gas with temperatures $k\Te \gtrsim 10\,{\rm keV}$ the convergence of the asymptotic expansion given above becomes slow \changeJ{(see Fig.~\ref{fig:precision_Itoh_CNSN})}. However, recently CNSN found an alternative set of frequency-dependent basis functions that allow very accurate description of the SZ signal up to high temperatures and bulk velocities.
\changeJ{The convergence radius of the CNSN expansion around the chosen pivot temperature (in CNSN $k T_{\rm e, 0} \simeq 15\,\keV$ was used) is also limited (see Fig.~\ref{fig:precision_Itoh_CNSN}), but in combination with the low temperature expansion it allows covering a large part of parameter space.}

Like for the asymptotic expansion, the signal is determined by particular temperature and velocity moments, but this time the weighting differs slightly from those of Eq.~\eqref{eq:SZ_signal_y_parameters}. 
With the expressions given in CNSN it is straightforward to show (see Appendix~\ref{app:der_new} for more details)  that up to some specified order, $k^\ast_{\rm max}$, of the electron temperature one has
%--------------
\bsub
\label{eq:SZ_signal_functions_hot}
\beal
Y^{\rm high}(x, \Te) &= 
\sum_{k=0}^{k^\ast_{\rm max}} Y^{\rm high}_k(x) \, \mathcal{N}(\The)\, \The^k
\\
M^{\rm high}(x, \Te) &= \sum_{k=0}^{k^\ast_{\rm max}} M^{\rm high}_k(x) \,\mathcal{N}(\The)\, \The^k
\\
D^{\rm high}(x, \Te)&=\sum_{k=0}^{k^\ast_{\rm max}}  
D^{\rm high}_k(x) \,\mathcal{N}(\The)\, \The^k
\\
Q^{\rm high}(x, \Te)&=\sum_{k=0}^{k^\ast_{\rm max}} Q^{\rm high}_k(x)\, \mathcal{N}(\The)\, \The^k,
\end{align}
\esub
%--------------
with $\mathcal{N}(\The)=\frac{e^{-1/\theta_{\rm e}}}{K_2(1/\theta_{\rm e})\,\theta_{\rm e}}\approx \frac{4\pi}{(2\pi\The)^{3/2}}\left[1-\frac{15}{2}\The+\frac{345}{128}\The^2+\mathcal{O}(\The^3)\right]$.
Here, $K_2(x)$ denotes the modified Bessel functions of second kind.
The functions $Y^{\rm high}_k$, $M^{\rm high}_k$, $D^{\rm high}_k$, and $Q^{\rm high}_k$ are defined by Eq.~\eqref{eq:SZ_signal_functions_defs_new_explicit}.
All temperature-independent kinematic terms were already taken into account by Eq.~\eqref{eq:SZ_signal_functions}, so that they do not reappear here.
Also, in general $k^\ast_{\rm max}\neq k_{\rm max}$, although below we usually set \changeJ{$k^\ast_{\rm max}\equiv k_{\rm max}$}.

In analogy to the low temperature gas case we introduce the following line-of-sight temperature and velocity moments:
%--------------
\beal
\label{eq:SZ_signal_y_parameters_new}
z^{(k)}&= \int  \mathcal{N}(\The) \,\The^{k} \id \tau,
&
c^{(k)}_0&= \int \betacsq \mathcal{N}(\The) \,\The^{k} \id \tau,
\\\nonumber
c^{(k)}_1&= \int \betac P_1(\muc) \mathcal{N}(\The) \,\The^{k} \id \tau,
&c^{(k)}_2&= \int \betacsq P_2(\muc) \mathcal{N}(\The) \,\The^{k} \id \tau,
\end{align}
%--------------
where the integrals are carried out in the cluster frame, with the condition $T_{\rm e, low}\leq \Te\leq T_{\rm e, high}$. 
The optimal value for $T_{\rm e, high}$ depends on the required precision. Also, one can split the temperature range up into different \changeJ{parts}, each with their own set of basis functions \changeJ{defined on the intervals $\Te \in (T_{\rm e, high}^{i-1}, T_{\rm e, high}^{i}]$}, as will be
specified in Sect.~\ref{sec:numberofmoments}.
With this we define the moment vector
%--------------
\beal
\vek{m}^T_{\rm high}&=(\vek{z}^T, \vek{c}_1^T, \vek{c}_0^T, \vek{c}_2^T)
\nonumber\\
\vek{z}^T&=(z^{(0)}, ..., z^{(k^\ast_{\rm max})})
\\\nonumber
\vek{c}_i^T&=(c_i^{(0)}..., c_i^{(k^\ast_{\rm max})}).
\end{align}
%--------------
The signal vector for the contribution of the high temperature gas, $\vek{S}_{\rm high}$, is defined similar to $\vek{S}_{\rm low}$, and it can be obtained with
%--------------
\beal
\label{sec:Signal_eq_high}
\vek{S}_{\rm high}&=\matrixf{F}_{\rm high} \vek{m}_{\rm high},
\end{align}
%--------------
where each row of the matrix $\matrixf{F}_{\rm high}$ reads 
%--------------
\beal
(\matrixf{F}_{\rm high})_{i}
=
\changeC{I_{\rm o}\,x_i^3}\left(
\vek{Y}^T_{\rm high}, \vek{D}^T_{\rm high}, 
\vek{M}^T_{\rm high}, \vek{Q}^T_{\rm high}
\right),
\end{align}
%--------------
with $\vek{Y}^T_{\rm high}=(Y^{\rm high}_0, ..., Y^{\rm high}_{k^\ast_{\rm max}})$, $\vek{D}^T_{\rm high}= (D^{\rm high}_0, ..., D^{\rm low}_{k^\ast_{\rm max}})$, and so on, all as above.
Again the SZ signal caused by high temperature gas can be expressed as a simple matrix multiplication, once the moments are determined. 
\changeJ{This reduces the computational burden to calculation of the temperature-velocity moment vector which solely depends on the cluster atmosphere.}

\subsection{Total line-of-sight SZ signal}
With \changeJ{the definitions of the previous sections,} the total SZ signal is given by $\vek{S}=\vek{S}_{\rm low}+\vek{S}_{\rm high}$.
Introducing the total cluster temperature-velocity moment vector, $\vek{m}^T=(\vek{m}^T_{\rm low}, \vek{m}^T_{\rm high})$, and the \changeC{frequency-dependent} matrix, $\matrixf{F}=(\matrixf{F}_{\rm low},\matrixf{F}_{\rm high})$, one has
%--------------
\beal
\label{sec:Signal_eq_tot}
\vek{S}=\matrixf{F}\,\vek{m}.
\end{align}
%--------------
It is clear that the dimension of $\vek{S}$ defines the maximal number of moments that could possibly be deduced from the SZ data.
However, in the presence of noise, foregrounds, and correlations between the moments, one naturally has ${\rm dim}(\vek{m})<{\rm dim}(\vek{S})$.

It is furthermore important that for a given experimental precision, an optimal combination of the basis functions can be found which minimizes the number of moments required to accurately represent the SZ signal.
In particular, with the approach of CNSN one can vary the \changeJ{reference/pivot} temperature, $T_{\rm e,0}$, and number of reference points used in the computation of the basis functions to improve the temperature coverage of the approximation.
In that case the moment vector can be cast into the form $\vek{m}^T=(\vek{m}^T_{\rm low}, \vek{m}^T_{\rm high I}, \vek{m}^T_{\rm high II}, ...)$ with different \changeJ{temperature regions, $[0, T_{\rm e, low}]$, $(T_{\rm e, low}, T_{\rm e, high}^{\rm I}]$, $(T_{\rm e, high}^{\rm I}, T_{\rm e, high}^{\rm II}]$, $(T_{\rm e, high}^{\rm II}, T_{\rm e, high}^{\rm III}]$, and so on}.
Here the subscript `low' is used to indicate that the asymptotic expansion is applied for those moments, while for any moments with subscript `high' we formulate the basis using CNSN.
We will discuss the associated optimization problem in Sect.~\ref{sec:numberofmoments}.

With the formulation given above it is also straightforward to include the effect of angular resolution and frequency bands on the SZ signal.
The effect of angular resolution is accounted for by spatially averaging the temperature-velocity moments, i.e., $\vek{m}\rightarrow \left<\vek{m}\right>$, where $\left<...\right>$ denotes angular/spatial average.
The bandpass can be taken into account by means of a matrix $\matrixf{W}$.
With this the SZ signal in more general can be expressed as
%--------------
\beal
\label{sec:av_Signal_eq_tot}
\vek{S}=\matrixf{W}\,\matrixf{F}\left<\vek{m}\right> + \vek{n},
\end{align}
%--------------
where we also added noise to the problem. In a similar way possible contaminations by (spatially) smooth foregrounds, radio sources, or dusty-star-forming galaxies (DSFGs) can be incorporated.

One of the benefits of the moment method described here is that for a given set of frequencies the moment matrix only has to be computed once. 
This, for example, makes computation of the SZ signal from simulated cluster very efficient and accurate.
However, Eq.~\eqref{sec:Signal_eq_tot} is still mainly interesting from the computational point of view because the entries of the moment vector are {\it not} independent. For instance, all moments related to $y^{(k)}$ are non-negative and one also expects $y^{(k)}>y^{(k+1)}$.
This imposes rather complicated priors and correlations among the different entries of the moment vector with the actual dimensionality of the problem being much smaller. We will show below that for future SZ observations only a few parameters are required to accurately describe the SZ signal, although the number of entries in the moment vector, $\vek{m}$, is much larger.

Finally, we mention that the effect of the observers motion with respect to the cluster \citep{Chluba2005b, ChlubaSZpack} can be included using simple Lorentz-transformation of the frequencies and corresponding angles into the CMB rest frame, to account for the effect of Doppler boosting and relativistic light aberration \citep[e.g.,][]{Chluba2011ab}. However, for the discussion below these aspects of the problem are not crucial and will be omitted.

%---------------
\begin{figure}
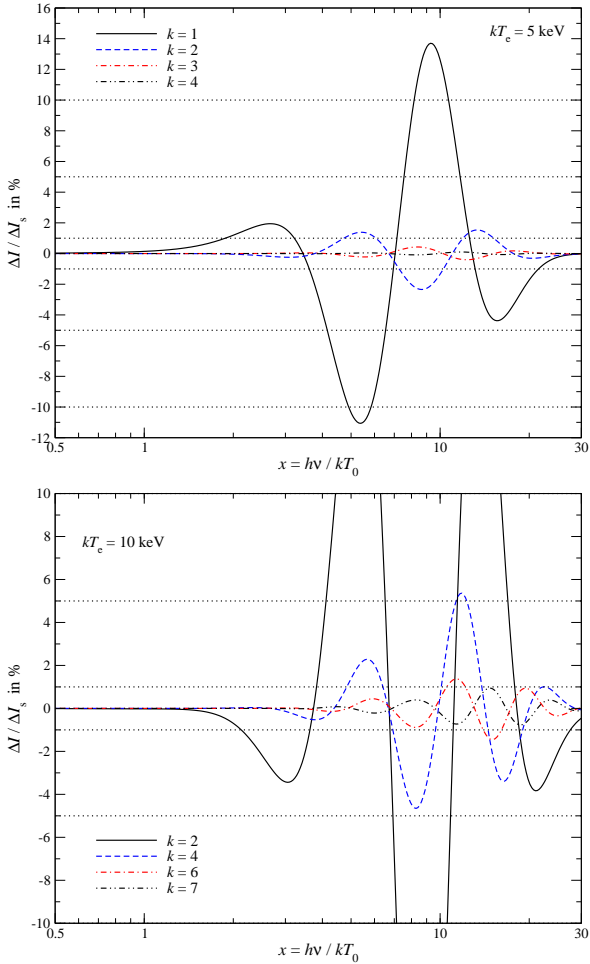

\centering
\includegraphics[width=0.92\columnwidth]{./eps/precision.Itoh.5keV.eps}
\\[1mm]
\includegraphics[width=0.92\columnwidth]{./eps/precision.Itoh.10keV.eps}
\caption{Deviation of the approximation from the numerical result. The asymptotic expansion is used to represent the SZ signal. For each of the curves, $k+1$ temperature orders were included. The departure is expressed in units of $\Delta I_{\rm s}\simeq 0.013\, {\rm MJy\,sr^{-1}}$ and $\tau=0.01$ was assumed.}
\label{fig:precision_Itoh}
\end{figure}
%---------------
%---------------
%---------------
\subsection{Minimizing the required number of \changeJ{moments}}
\label{sec:numberofmoments}
%------------------------------------
One of the important questions is how many moments are needed to describe the SZ signal accurately for a given experimental sensitivity and range of gas temperatures.
Here it is particularly interesting to try minimizing the total number of moments that are required to achieve an optimal representation of the SZ signal.
To answer this question we first define a fiducial sensitivity for comparison. We shall use the kSZ signal of a cluster with line-of-sight optical depth $\tau=10^{-2}$ and $\betac \muc=10^{-3}$ close to the thSZ crossover frequency $\nu_{\rm c} = 217\,{\rm GHz} \,(x_{\rm c}=3.83)$ as benchmark; this gives a distortion with amplitude $\Delta I_{\rm s}\simeq 9.76\,\tau\, \betac\muc (kT_0)^3 / [ h^2 c^2 ] \,{\rm sr^{-1}}
%\simeq \pot{1.32}{-19}\,{\rm ergs \,cm^{-2} s^{-1} Hz^{-1} sr^{-1}}
\simeq 0.013\, {\rm MJy\,sr^{-1}}$.
For an isothermal cluster with $k\Te \simeq 5\,{\rm keV}$ electrons and $\tau=10^{-2}$ the maximal thSZ signal is roughly $\Delta I_{\rm th}\simeq 0.17\, {\rm MJy\,sr^{-1}}$ at $x\simeq 6.7$.
Therefore, $\Delta I_{\rm s}$ corresponds to $\simeq 8\%$ precision on $\Delta I_{\rm th}$.

Before carrying out additional computations we extended the basis of CNSN with additional pivot points (at $\theta_{\rm e, 0}=0.01$ and $0.1$) such that the SZ signal can be represented in a wider range of temperatures. 
We provide this basis both in the cluster rest frame and the CMB frame.
With the current version of {\sc SZpack} a $\simeq 0.001\%$ precision is achieved at frequencies $0.01\lesssim x \lesssim 30$, temperatures $k\Te \lesssim 75\,{\rm keV}$, and for $\betac\lesssim 0.01$ basically at no additional computational cost.
We furthermore included the necessary database directly into {\sc SZpack} such that no time is consumed loading data. 
In the current implementation evaluation of the SZ signal at 400 frequencies takes about 0.01 seconds on a standard laptop. 
These routines can also be directly invoked from Python.

One can now calculate how accurately the different sets of basis functions describe the SZ signal for varying $k_{\rm max}$.
For the asymptotic expansion we show two examples in Fig.~\ref{fig:precision_Itoh}. At low temperatures (upper panel) only a few terms in the expansion are needed to achieve a very accurate representation of the SZ signal. The mismatch is usually largest at high frequencies, $x\simeq 10$, while below the crossover frequency higher order temperature terms are small, even for larger electron temperatures.
The lower panel of Fig.~\ref{fig:precision_Itoh} indicates that at higher temperatures the convergence of the asymptotic expansion becomes slower, a problem that is well-known from previous analysis \citep[e.g., see][]{Itoh98}.

A simple calculation can be used to further quantify the convergence rate of the different basis functions:
for a given order in temperature we compute the maximal deviation of the approximation from the numerical result in the frequency range $0.1\leq x \leq 30$.
For both the asymptotic expansion and the basis CNSN with reference temperature $\theta_{\rm e, 0}=0.03$ the results are shown in Fig.~\ref{fig:precision_Itoh_CNSN}. One can see that for the asymptotic expansion the agreement with the numerical result does not improve above $k\Te\simeq 13 \,{\rm keV}$. 
At higher temperatures the expansion of CNSN performs much better.
In particular, the convergence radius increases strongly when including the first few temperature corrections.
One can also observe that for the CNSN basis functions with pivot temperature $\theta_{\rm e, 0}=0.03$ convergence above $k\Te\simeq 40 \,{\rm keV}$ is not achieved. However, this can be overcome by adding another set of basis functions with pivot $\theta_{\rm e, 0}>0.03$.
For fixed temperature correction order, $k_{\rm max}$, one can therefore try to find an optimal combination of pivot temperatures to cover a large range of temperatures.
This is not the absolute minimum with respect to the number of moments, but optimization in this way still is beneficial while remaining sufficiently simple.

%---------------
\begin{table}
\caption{Optimal distribution of temperature pivots, \changeJ{$T^i_{\rm e, 0}$, and regions, $[0, T_{\rm e, low}]$, $(T_{\rm e, low}, T_{\rm e, high}^{\rm I}]$, $(T_{\rm e, high}^{\rm I}, T_{\rm e, high}^{\rm II}]$, etc,} for given accuracy goal, \changeJ{$\Delta I/\Delta I_{\rm s}$. At} temperature $\Te\leq T_{\rm e, low}$ the asymptotic expansion is used, while above expressions based on CNSN are applied.
As fiducial accuracy value we used $\Delta I_{\rm s}\simeq 0.013\, {\rm MJy\,sr^{-1}}$ and optical depth $\tau=0.01$. Numbers marked with asterisk are only estimated upper bounds, although the approximation is much better. \changeJ{The number of temperature correction terms in each region is $k_{\rm max}$, while $k_{\rm tot}$ gives the total number of temperature terms for all regions. The required number of temperature-velocity moments depends on the settings for kinematic corrections and is not further specified here.}}
\label{tab:convergence}
\centering
\begin{tabular}{@{}ccccccc}
\hline
\hline
$\Delta I/\Delta I_{\rm s}$ & $k_{\rm max}/k_{\rm tot}$ 
& $T_{\rm e, low}$ 
& $T_{\rm e, high}^{\rm I}/T^{\rm I}_{\rm e, 0}$ 
& $T_{\rm e, high}^{\rm II}/T^{\rm II}_{\rm e, 0}$ 
& $T_{\rm e, high}^{\rm III}/T^{\rm III}_{\rm e, 0}$ 
\\
&  & $[\rm keV]$ & $[\rm keV]$ & $[\rm keV]$ & $[\rm keV]$  \\
\hline
1 & 2/12 & 9.1 & $21.6 / 14$ & $42 / 30$ & $75 / 55$ \\
1 & 3/12 & 12.5 & $48.5 / 25$ & $90^\ast / 80$ & -- \\
1 & 4/10 & 14.3 & $75 / 35$ & -- & -- \\
%1 & 3/8 & 15.3 & $54 / 30$ & -- & -- \\
\hline
0.1 & 3/16 & 6.8 & $17.25 / 11$ & $36.4 / 25$ & $68 / 50$ \\
0.1 & 4/15 & 9.3 & $33.1 / 18.5$ & $80^\ast / 55$ & -- \\
0.1 & 5/18 & 10.76 & $48.5 / 25$ & $90^\ast / 80$ & -- \\
\hline
0.01 & 4/20 & 5.76 & $14.7 / 9.4$ & $32 / 22$ & $61 / 45$ \\
0.01 & 5/18 & 7.3 & $23.86 / 14$ & $61 / 40$ & -- \\
0.01 & 6/21 & 8.55 & $33.3 / 18.5$ & $80^\ast / 60$ & -- \\
\hline
$\pot{5}{-4}$ & 6/28 & 5.76 & $14.7 / 9.4$ & $32 / 22$ & $63.5 / 45$ \\
%0.001 & 6/18 & 7.3 & $23.86 / 14$ & $61 / 40$ & -- \\
%0.001 & 7/21 & 8.55 & $33.3 / 18.5$ & $80^\ast / 60$ & -- \\
%
\hline
\hline
\end{tabular}
\end{table}
%---------------

In Table~\ref{tab:convergence} we summarize the results of our efforts to cover at least the temperature range $0\leq k\Te \lesssim 60\,{\rm keV}$ for a given precision and $k_{\rm max}$. 
We defined different regions of temperatures making sure that close to the boundaries the condition on the precision is met with some $10\%-20\%$ margin.
Far away from the boundaries of the different temperature regions the approximations are typically much more accurate.
The setting for accuracy goal $\pot{5}{-4}\Delta I_{\rm s}$ is already close to the numerical precision of our approximations and is mainly meant to provide an extreme setting for comparisons.
Furthermore, if the electron temperature is smaller than some maximal temperature, $T_{\rm e, max}$, the total number of required variables can be further reduced by dropping moments in regions with $\Te > T_{\rm e, max}$. 
For a given accuracy goal this defines an optimal value for $k_{\rm max}$. 
We also found that the same settings work when $0\lesssim\betac \lesssim 0.01$.

One point we mention is that the settings given in Table~\ref{tab:convergence} are in fact independent of the chosen optical depth, $\tau$.
This means that only the scaling of the approximation with electron temperature affects the precision. For instance, if the optical depth along a given line-of-sight is $\tau\simeq \pot{2}{-3}$, but the temperature is fixed, then the absolute precision of the approximation at accuracy goal in the second category (denoted with $0.1 \Delta I_{\rm s}$) is actually $\lesssim 0.02 \Delta I_{\rm s}$.

\section{SZ signals for various cluster models}
\label{sec:SZclusters}
%------------------------------------
To demonstrate how to use and interpret the temperature-velocity moment method we 
now discuss the SZ signals for different cluster models. We start with the 
simplest case of an isothermal cluster and then work our way through 
several instructive examples, also introducing the simpler moment method that is applicable to sufficiently \changeJ{smooth (low temperature-velocity variance) cluster atmospheres}.

\subsection{SZ signals for isothermal clusters}
\label{sec:isothermal_clusters}
%------------------------------------
Traditionally, galaxy clusters have been modelled using a simple isothermal $\beta$-model \citep{Cavaliere1978}:
%--------------
\beal
\label{sec:Ne_prof_iso}
%\Ne&=\frac{N_{\rm e,0}}{[1+(r/r_{\rm c})^2]^{3\beta/2}}
\Ne&=N_{\rm e,0}[1+(r/r_{\rm c})^2]^{-3\beta/2}
\quad{\rm and} %\nonumber\\
\quad
\Te\equiv {\rm const},
\end{align}
%--------------
where $N_{\rm e,0}\simeq 10^{-3} {\rm cm^{-3}}$ is the typical central number density of free electrons, $r_{\rm c}\simeq 100\,{\rm kpc}$ is the typical core radius of clusters, and $\beta \simeq 2/3$ \citep[e.g., see][]{Reese2002}.
In the absence of bulk velocities, one therefore has the temperature moments $y^{(k)}=\The^{k+1}\tau$ for $\Te\leq T_{\rm e, low}$ and $z^{(k)}=\mathcal{N}(\The)\,\The^{k}\tau$ for $\Te>T_{\rm e, low}$.
This shows that the spatial morphology of the SZ signal is fully determined by the overall optical depth factor, $\tau(\vgh)$ \citep[we shall ignore small corrections caused by multiple scattering, e.g., see][]{Dolgov2001, Itoh2001, Colafrancesco2003}, with the same spectral shape for any line-of-sight through the cluster.
This also implies that the average SZ signal measured for an unresolved cluster in this case is determined by only one spectral function, and the spatially averaged optical depth, $\left<\tau\right>$.

Allowing the cluster to move with a peculiar velocity, $\vbetac$, relative to the CMB one readily obtains the velocity moments\footnote{We neglected tiny corrections caused by deviations from the flat-sky approximation \citep{ChlubaSZpack}.} $b^{(k)}_0=\betac^2 \The^{k}\tau$, $b^{(k)}_1\approx \betac \muc \The^{k}\tau$ and $b^{(k)}_2 \approx \betac^2 P_2(\muc) \The^{k}\tau$ for clusters with $\Te\leq T_{\rm e, low}$, and $c^{(k)}_0=\betac^2\mathcal{N}(\The)\,\The^{k}\tau$, $c^{(k)}_1 \approx \betac\muc \mathcal{N}(\The)\,\The^{k}\tau$, and $c^{(k)}_2 \approx\betac^2 P_2(\muc) \mathcal{N}(\The)\,\The^{k}\tau$ for hot clusters.
Again the spatial dependence of the SZ signal factors out and is determined by the one of the line-of-sight optical depth alone.
It is however clear that the SZ morphology becomes frequency-dependent\footnote{Here we mean that the spectral dependence of the SZ signal no longer is independent of the line-of-sight.} once the cluster no longer is isothermal or significant internal motions of the ICM are present.
The reason for this frequency-dependence is related to the variation of the temperature-velocity moments along different lines-of-sight, as we illustrate below (Sect.~\ref{sec:cluster_sims}).

\subsection{SZ signals for smooth density and temperature profiles}
\label{sec:example_clusters}
%------------------------------------
In more realistic cluster models, the variation of the temperature also has to be included. 
One common possibility assumes a polytropic temperature profile, $\Te \propto \rho_{\rm gas}^{1-\gamma}$ \citep{Markevitch1999, Finoguenov2001, Pratt2002}.
Alternatively, one can consider fits to the observed temperature and density profiles derived from {\it Chandra} X-ray data \citep{Vikhlinin2006}.

For our discussion it is only important that the associated profiles are very smooth. This suggests that a good approximation for the SZ signal can be found by computing average values for the temperature and velocity along the line-of-sight.
Corrections to this lowest order approximation can then be included using a Taylor-series around the average values.
Assuming that $\betac=0$, we can introduce the SZ-weighted electron temperature, 
%--------------
\beal
\label{sec:TSZ_prof}
T_{\rm e, SZ}(\vgh)&=(\me c^2/k) \, y^{(0)}/y^{(-1)}=\tau^{-1}\!\int \Te \id \tau.
\end{align}
%--------------
The integrals $y^{(k)}$ are defined by Eq.~\eqref{eq:SZ_signal_y_parameters} but here we do not impose any condition on the electron temperature.
One can furthermore introduce the isothermal temperature moments 
%--------------
\beal
\label{sec:yISO_prof}
y_{\rm iso}^{(k)}(\vgh)=[kT_{\rm e, SZ}(\vgh)/\me c^2]^{k+1}\, \tau(\vgh).
\end{align}
%--------------
%
In general these moments are {\it not} identical to $y^{(k)}(\vgh)$ for $k>0$, and the ratio $\rho^{(k)}(\vgh)=y^{(k)}/y_{\rm iso}^{(k)}\equiv \tau^k y^{(k)}/ [y^{(0)}]^{k+1}$
can be used to quantify departures from isothermality:
in regions with $\rho^{(k)} \neq 1$ one expects the SZ signal to be poorly represented by just using the SZ-weighted electron temperature and line-of-sight optical depth.
The ratios $\rho^{(k)}$ are also independent of the overall temperature and density scales. They only depend on the {\it shapes} of the cluster temperature and electron density profiles.

We can write this more formally by using the isothermal moment vector, $\vek{m}_{\rm iso}(\tau, T_{\rm e, SZ})$, and expanding the average SZ signal around $\TeSZ$ and $\tau$:
%--------------
\beal
%\label{sec:Signal_eq_tot_series_simp}
\vek{S}&\approx \matrixf{F}\,\left[\vek{m}_{\rm iso}(\tau, T_{\rm e, SZ})+\sum_{k=2}^{k_{\rm max}} \frac{\partial^k_{\theta_{\rm e, SZ}}\vek{m}_{\rm iso}(T_{\rm e, SZ})}{ \tau\, k!}\int(\The-\theta_{\rm e, SZ})^k\id\tau\right]
\nonumber
\end{align}
%--------------
where the first order derivative term canceled after performing the line-of-sight average.
Here $\vek{S}_{\rm iso}(\tau, T_{\rm e, SZ})=\matrixf{F}\,\vek{m}_{\rm iso}(\tau, T_{\rm e, SZ})$ is the leading order, average SZ signal, while $\partial^k_{\The}\vek{S}_{\rm iso}(\Te)=\matrixf{F}\,\partial^k_{\The}\vek{m}_{\rm iso}(\Te)$ is the derivatives of the SZ signal with respect to $\The$.
To simplify the notation, we furthermore introduce 
%--------------
\beal
%\label{sec:Dy_def}
\Delta y^{(k)}&=\int (\The-\theta_{\rm e, SZ})^{k+1} \id\tau
=y^{(k)}_{\rm iso}\sum_{m=0}^{k+1}\binom{k+1}{m}\,(-1)^{k+1-m} \rho^{(m-1)}.
\nonumber
\end{align}
%--------------
This means $\Delta y^{(0)}=0$, $\Delta y^{(1)}=y^{(1)}-y^{(1)}_{\rm iso}=y^{(1)}_{\rm iso}\,\Delta \rho^{(1)}$, $\Delta y^{(2)}=y^{(2)}_{\rm iso}\,\Delta\rho^{(2)}-3y^{(2)}_{\rm iso}\,\Delta\rho^{(1)}$, and so on, with $\Delta\rho^{(k)}=\rho^{(k)}-1$.
Defining $\omega^{(k)}=\Delta y^{(k)}/y^{(k)}_{\rm iso}\equiv \tau^k \Delta y^{(k)}/[y^{(0)}]^{k+1}$ we then can finally write
%--------------
\beal
\label{sec:Signal_eq_tot_series_simpfinal}
\vek{S}&\approx
\vek{S}^{(0)}_{\rm iso}(\tau, T_{\rm e, SZ})
+\vek{S}^{(2)}_{\rm iso}(\tau, T_{\rm e, SZ})\,\omega^{(1)}
+\vek{S}^{(3)}_{\rm iso}(\tau, T_{\rm e, SZ})\,\omega^{(2)}
+ ...
\end{align}
%--------------
where \changeC{$\vek{S}^{(k)}_{\rm iso}(\tau, T)=(T^{k}/k!)\,\partial^{k}_{T}\vek{S}_{\rm iso}(\tau,T)$}. 
In this parametrization the observables for the SZ measurement are $\tau$, $\TeSZ$, and the temperature moments $\omega^{(k)}$.

\begin{figure}
\centering
\includegraphics[width=0.92\columnwidth]{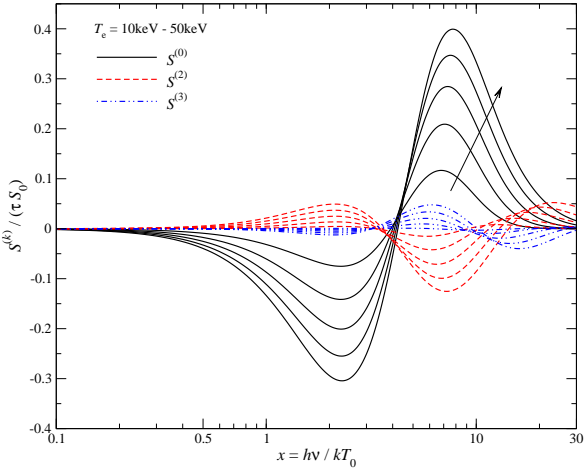}
\caption{Spectral functions $\vek{S}^{(k)}_{\rm iso}$ in units of $\tau$ and $S_0=(2h/c^2) (kT_0/h)^3\approx 270\,\MJysr$. For comparison we also show the main SZ signal (solid black line). The arrow indicates the direction of increasing temperature, with the lines within the groups being separated by $\Delta \Te=10\,{\rm keV}$.}
\label{fig:SZ_derivs}
\end{figure}

With {\sc SZpack} it is straightforward to compute the required vectors, $\vek{S}^{(k)}_{\rm iso}$, with very high precision.
In Fig.~\ref{fig:SZ_derivs} we show the first few $\vek{S}^{(k)}_{\rm iso}$.
%---------------
%---------------
The typical amplitude of the $\vek{S}^{(k)}_{\rm iso}$ is dropping with $k$, indicating that unless the temperature moments $\omega^{(k)}$ increase strongly with $k$, higher order terms remain small with the largest contributions at high frequencies.
This shows that the considered expansion becomes fully perturbative unless rather large deviations from the smooth temperature profile case are present.

For example, using the simple fits for one of the hottest clusters ($\Te\simeq 9.2\,\keV$ at $r\simeq 330\,\kpc$), A2029, from the cluster sample of \citet{Vikhlinin2006}, we find $\omega^{(1)}\simeq 0.16$, $\omega^{(2)}\simeq 0$, and $\omega^{(3)}\simeq 0.05$ close to the cluster center.
This indicates that higher order corrections decay rapidly.
In fact, the correction related to $\omega^{(1)}$ contributes at the level of a few percent to the average SZ signal, while higher order moments are negligible.
We find that even for more realistic cases from simulated clusters only a few moments of the temperature field need to be known to accurately describe the SZ signal (Sect.~\ref{sec:cluster_sims}).
Also, in the more extreme case of a two-temperature plasma only the first few terms are required (Sect.~\ref{sec:example_two_temp_cluster}).

Some interesting frequencies for the functions $\vek{S}^{(k)}_{\rm iso}$ are related to \changeJ{their} nulls, maxima and minima. For $\vek{S}^{(1)}_{\rm iso}$ (which is not shown in Fig.~\ref{fig:SZ_derivs}) we find a rather temperature-independent minimum at $x\simeq 2.26$. At $x\simeq 4$ it crosses zero for $\Te\simeq 10\,\keV$, while for $\Te\simeq 50\,\keV$ the null is located at $x\simeq 4.89$. Its maximum is located at $x\simeq 7.2$ for $\Te=10\,\keV$ and at $x\simeq 10$ for $\Te=50\,\keV$. 
For $\vek{S}^{(2)}_{\rm iso}$ (see Fig.~\ref{fig:SZ_derivs}) on the other hand we find a maximum at $x\simeq 2.12$ and the first null at $x\simeq 3.5$. 
The position of the minimum varies from $x\simeq 5.8$ for $\Te=10\,\keV$ to $x\simeq 7.2$ for $\Te=50\,\keV$.
These properties might be useful when deciding about the locations of frequency channels in future SZ experiments.

We emphasize that for the computation of $\vek{S}^{(0)}_{\rm iso}(\tau, T_{\rm e, SZ})$ and $\vek{S}^{(2)}_{\rm iso}(\tau, T_{\rm e, SZ})$ a large number of temperature terms has to be included. 
Although in the example given above we only find $\omega^{(1)}$ to be significant as additional parameter, this is not equivalent to dropping higher order temperature terms.
This point is very important when interpreting future SZ data, since otherwise biased results for $\tau, T_{\rm e, SZ}$, and $\omega^{(1)}$ are obtained.
Similarly, one has to include $\betac\muc \neq 0$ for the analysis, as we discuss in more detail below.

\changeJ{We also note that Eq.~\eqref{sec:Signal_eq_tot_series_simpfinal} is applicable even if the real temperature distribution is not a smooth function. It is only important that the variance of the temperature and thermal pressure remains sufficiently low to warrant decreasing values of the moments, $\omega^{(k)}$, with larger $k$. This condition is usually fulfilled even in more realistic cluster models (see Sect.~\ref{sec:cluster_sims}).}

\subsubsection{The effect of line-of-sight temperature variance}
\label{sec:smooth_SZ_omega}
Above we showed that the dominant correction to the SZ signal is determined by the temperature moment $\omega^{(1)}$. This parameter can be interpreted as line-of-sight variance or dispersion of the electron temperature but weighted by the optical depth of the scattering volume element.
One can now address the question of how important this term is for the interpretation of the SZ signal. In particular, by how much are the deduced best-fit values for $\TeSZ$ and $\tau$ affected if the contribution from $\omega^{(1)}$ is neglected for high-sensitivity, multi-frequency SZ measurements.

\begin{figure}
\centering
\includegraphics[width=0.92\columnwidth]{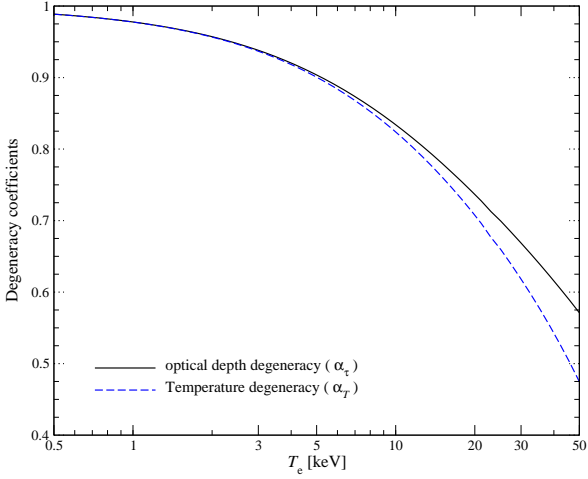}
\caption{Degeneracies of $\tau$ and $T_{\rm e, SZ}$ with $\omega^{(1)}$. The expected (biased) best-fit values are $\tau^\ast \approx\tau[1-\alpha_{\tau}(\Te)\,\omega^{(1)}]$ and $T_{\rm e, SZ}^\ast\approx T_{\rm e, SZ}[1+\alpha_{T}(\Te)\,\omega^{(1)}]$ when analyzing the SZ signal.}
\label{fig:degen_coeffy}
\end{figure}

We can start by writing the SZ signal for $T^\ast_{\rm e, SZ}\neq T_{\rm e, SZ}$ as an expansion of $\vek{S}(\tau^\ast, T^\ast_{\rm e, SZ})$ around $T_{\rm e, SZ}$ and $\tau$:
%
%--------------
\beal
%\label{sec:Signal_shifted_TSZ}
\vek{S}(\tau^\ast, T^\ast_{\rm e, SZ})&\approx
\vek{S}^{(0)}_{\rm iso}(\tau, T_{\rm e, SZ})
+\changeC{\vek{S}^{(0)}_{\rm iso}}(\tau, T_{\rm e, SZ})\,\Theta_\tau
+\vek{S}^{(1)}_{\rm iso}(\tau, T_{\rm e, SZ})\,\Theta,
\nonumber
%+\vek{S}(\tau, T_{\rm e, SZ})\Delta\tau
%\nonumber\\
%&\qquad
%+\vek{S}^{(1)}_{\rm iso}(\tau, T_{\rm e, SZ})\,\Theta_\tau\,\Theta
%+\vek{S}^{(2)}_{\rm iso}(\tau, T_{\rm e, SZ})\,\Theta^2
\end{align}
%--------------
where $\Theta=(T^\ast_{\rm e, SZ}-T_{\rm e, SZ})/T_{\rm e, SZ}$, $\Theta_\tau=(\tau^\ast-\tau)/\tau$, and we neglected higher order terms.
To determine the best-fit values for $\tau^\ast$ and $T^\ast_{\rm e, SZ}$ one has to compare to $\vek{S}(\tau, T_{\rm e, SZ}, \omega^{(1)})\approx
\vek{S}^{(0)}_{\rm iso}(\tau, T_{\rm e, SZ})+\vek{S}^{(2)}_{\rm iso}(\tau, T_{\rm e, SZ})\,\omega^{(1)}$ and then minimize the squared difference.
The coefficients relating $\Theta_\tau$ and $\Theta $ to $\omega^{(1)}$ then are only functions of temperature, and it is straightforward to compute the degeneracy coefficients $\alpha_{\rm \tau}=-\Theta_\tau/\omega^{(1)}$ and $\alpha_{T}=\Theta/\omega^{(1)}$ (Fig.~\ref{fig:degen_coeffy}). 
Both $\tau$ and $\TeSZ$ are correlated with $\omega^{(1)}$ to a similar degree.
The degeneracy is very close to unity at low temperatures and only drops to about $1/2$ at very high temperatures. 
We find that $\alpha_{\tau}(\Te)\approx [1+\pot{2.7}{-2}\,\Te^{0.86}]^{-1}$ and $\alpha_{T}\approx \exp\left(-\pot{2.6}{-2}\,\Te^{0.86}\right)$ match the full numerical result with $\simeq 10\%$ precision.
With these expressions we can directly estimate the expected value for $T^\ast_{\rm e, SZ}$ obtained by computing the best-fits to the full SZ signal.
%

%---------------
%---------------
To \changeJ{determine} the degeneracy coefficients we used a very dense grid of frequency points in the range $x=0.1$ to $30$. More realistically far fewer independent frequency bins are available plus the signal is averaged over some bandwidth and the beam. Furthermore, foregrounds and the experimental sensitivity at each frequency are important.
All these aspects affect the degeneracy between the SZ parameters, as we explain in more detail below (Sect.~\ref{sec:mock}).
Nevertheless, the estimate obtained above gives a rough scaling for the importance of line-of-sight temperature variations for the interpretation of the SZ measurement.

\subsubsection{The effect of velocity terms on the SZ signal}
\label{sec:smooth_SZ_beta}
Thus far we neglected the effect of bulk velocity on the SZ signal. However, the effect of (internal) motions can again be included by expanding the SZ signal around the mean line-of-sight values.
We first define the two velocity components $\beta_{\rm c, \parallel}=\betac\muc$ and $\beta_{\rm c, \perp}=\betac\sqrt{1-\muc^2}$ of the moving volume element. Since we do not consider the effect of polarization these are the only two variables that matter for the SZ signal.
At lowest order in $\betac$ one is only sensitive to the line-of-sight SZ weighted averaged velocity $\beta_{\rm c, \parallel, SZ}=\tau^{-1}\int \beta_{\rm c, \parallel} \id \tau$. 
The associated kSZ signal has a spectral shape that is very different from the thSZ. It is therefore rather easy to distinguish the two. 
However, the results for the best-fit values of $\tau$ and $\TeSZ$ can be strongly biased even for a small ($\betacpara\simeq 10^{-3}$) line-of-sight velocity \changeJ{if neglected in the analysis} (see Sect.~\ref{sec:mock}).

Looking at the total SZ signal, Eq.~\eqref{eq:SZ_signal}, with $\betac^2 = \beta^2_{\rm c,\parallel}+\beta^2_{\rm c, \perp}$ and $\betac^2 P_2(\muc)=\beta^2_{\rm c, \parallel}-\beta^2_{\rm c, \perp}$, it is clear that the SZ signal is only sensitive to $\beta^2_{\rm c, \perp, SZ}=\tau^{-1}\int \beta^2_{\rm c, \perp} \id \tau$, but not $\beta_{\rm c, \perp, SZ}$ directly.
For simplicity (and because the associated terms are extremely small) we shall neglect the variation of temperature-velocity cross terms of order $\mathcal{O}(\betac^2\The^k)$ in the discussion here. 
Expanding around $\tau$, $T_{\rm e, SZ}$, $\beta_{\rm c, \parallel, SZ}$, and $\beta_{\rm c, \perp, SZ}=0$, the average SZ signal takes the form
%--------------
\beal
\label{sec:Signal_eq_tot_series_velocity}
\vek{S}&\approx
\vek{S}^{(0)}_{\rm iso}
+\vek{S}^{(2)}_{\rm iso}\,\omega^{(1)}
+\vek{C}^{(1)}_{\rm iso}\,\sigma^{(1)}
+\vek{D}^{(2)}_{\rm iso}\,\kappa^{(1)}
+\vek{E}^{(2)}_{\rm iso}\,\beta^2_{\rm c, \perp, SZ}
+ ...,
\end{align}
%--------------
where we suppressed the function arguments $\tau$, $T_{\rm e, SZ}$, and $\beta_{\rm c, \parallel, SZ}$.
We furthermore defined the signal vectors and dispersion variables,
%--------------
\beal
\label{sec:Signal_eq_tot_series_velocity_functions}
\vek{C}^{(k)}_{\rm iso}&=
(k!\,m!)^{-1}\,\theta^{k}_{\rm e, SZ}
\partial^{k}_{\theta_{\rm e, SZ}} 
\partial_{\beta_{\rm c, \parallel, SZ}} 
\vek{S}_{\rm iso}(\tau, T_{\rm e, SZ}, \beta_{\rm c, \parallel, SZ}, \beta_{\rm c, \perp, SZ})
\nonumber
\\
\vek{D}^{(k)}_{\rm iso}&=(k!)^{-1}\,\partial^{k}_{\beta_{\rm c, \parallel, SZ}} \vek{S}_{\rm iso}(\tau, T_{\rm e, SZ}, \beta_{\rm c, \parallel, SZ}, \beta_{\rm c, \perp, SZ})
\nonumber
\\
\vek{E}^{(k)}_{\rm iso}&=(k!)^{-1}\,\partial^{k}_{\beta_{\rm c, \perp, SZ}} \vek{S}_{\rm iso}(\tau, T_{\rm e, SZ}, \beta_{\rm c, \parallel, SZ}, \beta_{\rm c, \perp, SZ})
\nonumber
\\
\kappa^{(k)}&=\tau^{-1}
\int (\beta_{\rm c, \parallel}-\beta_{\rm c, \parallel, SZ})^{k+1} \id\tau
\nonumber
\\
\sigma^{(k)}&=
(T^k_{\rm e, SZ}\tau)^{-1}
\int (\Te-T_{\rm e, SZ})^k(\beta_{\rm c, \parallel}-\beta_{\rm c, \parallel, SZ}) \id\tau,
\end{align}
%--------------
where the derivatives are evaluated at $\beta_{\rm c, \perp, SZ}=0$. Since in general $\betac$ can vanish, we did not rescale the velocity derivative terms, and correspondingly defined the velocity moments without weighting by the mean values.

As we argue below, for future multi-frequency SZ analysis the average SZ signal is described very well by Eq.~\eqref{sec:Signal_eq_tot_series_velocity}. 
This implies that the SZ parameter estimation problem for a single lines-of-sight boils down to constraining the set of SZ parameters $\vek{p}=(\tau, \TeSZ, \beta_{\rm c, \parallel, SZ}, \omega^{(1)}, \sigma^{(1)}, \kappa^{(1)}, \beta^2_{\rm c, \perp, SZ})$.
Here we ordered the parameters according to their expected importance for the average SZ signal.
Higher order temperature and velocity moments are straightforward to add to the problem and maps of the deduced quantities then allow constraining the structure of the cluster.
Maps of the different components of $\vek{p}$ therefore provide a compressed version of future multi-frequency, high resolution SZ data cubes. These new variables then directly allow constraining properties of the \changeJ{3-dimensional} cluster's atmosphere.

%---------------
\begin{figure}
\centering
\includegraphics[width=0.92\columnwidth]{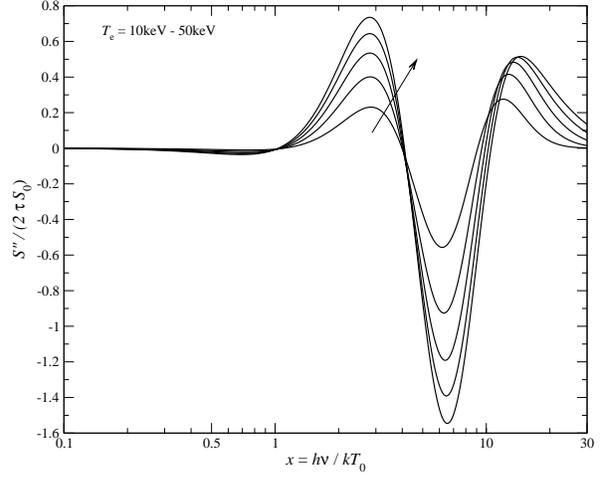}
\caption{Second frequency derivative of thSZ distortion in units of $\tau$ and $S_0=(2h/c^2) (kT_0/h)^3\approx 270\,\MJysr$. The arrow indicates the direction of increasing temperature, with the lines being separated by $\Delta \Te=10\,{\rm keV}$.}
\label{fig:S_xx}
\end{figure}
%---------------

\subsubsection{Effect of angular resolution on the SZ signal}
\label{sec:angular_res}
%----------------------
When deducing the best-fit parameters for $\tau$, $T_{\rm e, SZ}$, and $\beta_{\rm c, \parallel, SZ}$ it is also important to include the effect of angular resolution on the measured SZ signal. Gradients of the temperature and velocity field within the beam affect these values, as we illustrate here.

One can model the beam as a simple 2-dimensional Gaussian with radius, $r_{\rm res}$, or a beam variance \changeJ{$\sigma_{\rm b}^2=r_{\rm res}^2/[2\ln(2)] \approx 0.721\, r^2_{\rm res}$}. 
With the moment method described above we simply have to replace $\vek{m}$ with $\left<\vek{m}\right>$ in Eq.~\eqref{sec:Signal_eq_tot}, where $\left<...\right>$ denotes the average over the beam.
For a given electron density, temperature and velocity field, the associated 3-dimensional integrals are in principle straightforward to compute. However, in detail this can already be rather demanding, since several integrals have to be evaluated, respecting the conditions for the different temperature regions.

On the other hand, assuming rather smooth cluster profiles and small beam radius the calculation again simplifies significantly. Let the beam be centered on $\vek{r}_0=(r_{x, 0}, r_{y, 0})$ with central values $\tau_0$, $T_{\rm e, SZ, 0}$, and $\beta_{\rm c, \parallel, SZ, 0}$.
The average SZ signal around $\vek{r}_0$ can be computed from Eq.~\eqref{sec:Signal_eq_tot_series_velocity} and \eqref{sec:Signal_eq_tot_series_velocity_functions} by replacing $\tau\rightarrow \left<\tau\right>$, $T_{\rm e, SZ} \rightarrow \left<T_{\rm e, SZ}\right>$, and $\beta_{\rm c, \parallel, SZ} \rightarrow \left<\beta_{\rm c, \parallel, SZ}\right>$. Furthermore, one has to substitute $\beta^2_{\rm c, \perp, SZ}\rightarrow \beamav{\beta^2_{\rm c, \perp, SZ}}\approx \beta^2_{\rm c, \perp, SZ, 0}$ and compute the beam averages of $\omega^{(1)}$, $\sigma^{(1)}$, and $\kappa^{(1)}$, which read:
%--------------
\beal
\label{sec:beam_average_dispersions}
\beamav{\omega^{(1)}}&\approx \beamav{\tau}^{-1}\,\beamav{T_{\rm e, SZ}}^{-2}
\beamav{\int \left(\Te-\beamav{T_{\rm e, SZ}}\right)^2 \id\tau}
\nonumber
\\
\beamav{\kappa^{(1)}}& \approx \beamav{\tau}^{-1}
\beamav{\int \left(\beta_{\rm c, \parallel}-\beamav{\beta_{\rm c, \parallel, SZ}}\right)^2 \id\tau}
\\
\nonumber
\beamav{\sigma^{(1)}}&=
\beamav{\tau}^{-1}\beamav{T_{\rm e, SZ}}^{-1}
\beamav{\int \left(\Te-\beamav{T_{\rm e, SZ}}\right)
\left(\beta_{\rm c, \parallel}-\beamav{\beta_{\rm c, \parallel, SZ}}\right) \id\tau}.
\end{align}
%--------------
These dispersions have two main contributions, one from the variation along the line-of-sight and the other from the variation among different lines-of-sight inside the beam, which can be separated. For instance, by writing $\Te-\beamav{T_{\rm e, SZ}}=\Te-T_{\rm e, SZ}+T_{\rm e, SZ}-\beamav{T_{\rm e, SZ}}$, to second perturbation order we find 
%--------------
\beal
\label{sec:beam_average_dispersions_approx_II}
\beamav{\omega^{(1)}}&\approx \omega^{(1)}_0
+\beamav{T_{\rm e, SZ}}^{-2}\left(\beamav{T_{\rm e, SZ}^2}-\beamav{T_{\rm e, SZ}}^2\right)
\nonumber
\\
\beamav{\kappa^{(1)}}& \approx \kappa^{(1)}_0
+\beamav{\beta_{\rm c, \parallel, SZ}}^{-2}\left(\beamav{\beta^2_{\rm c, \parallel, SZ}}-\beamav{\beta_{\rm c, \parallel, SZ}}^2\right)
\\
\nonumber
\beamav{\sigma^{(1)}}&\approx 
\sigma^{(1)}_0+
\beamav{T_{\rm e, SZ}}^{-1}
\left(\beamav{T_{\rm e, SZ}\,\betacpar}-\beamav{T_{\rm e, SZ}}\beamav{\betacpar}\right),
\end{align}
%--------------
where, for example, $\omega^{(1)}_0=\omega^{(1)}(\vek{r}_0)$, is the line-of-sight temperature variance at the beam center.
Performing a Taylor expansion of quantities around $\vek{r}_0$ up to second order we have
%--------------
\beal
\label{sec:Taylor_beam}
\beamav{X}
&\approx X_0+\frac{1}{2}\,\sigma^2_{\rm b}\left[\partial^2_{r_x} + \partial^2_{r_y}\right] X_0 
\nonumber\\
\beamav{X^2}-\beamav{X}^2
&\approx \sigma^2_{\rm b} \left[(\partial_{r_x} X_0)^2 + (\partial_{r_y} X_0)^2 \right]
\\\nonumber
\beamav{X Y}-\beamav{X}\beamav{Y}
&\approx \sigma^2_{\rm b} \left[(\partial_{r_x} X_0)(\partial_{r_x} Y_0) + (\partial_{r_y}X_0)(\partial_{r_y}Y_0) \right]
\end{align}
%--------------
As these expressions show, the beam average values are affected only by terms related to the second derivatives, while the variances depend on first derivative terms.
Depending on the position of the beam with respect to the cluster center this introduces an interesting spatial dependence of $\beamav{\omega^{(1)}}, \beamav{\sigma^{(1)}}$, and $\beamav{\kappa^{(1)}}$. The precise values depend on the structure of the cluster, and only in the limit of very small beam radius will they be dominated by the line-of-sight variances. When deducing cluster parameters from SZ measurements this aspect has to be kept in mind. If this effect is ignored it can again lead to biases in the inferred parameters.

\subsubsection{Effect of frequency resolution on the SZ signal}
\label{sec:frequencies}
%--------------------------------
To estimate the effect of frequency resolution on the SZ signal we can use the fact that the SZ signal is a very smooth function of $x$. Defining the average frequency $\bar{x}$ and dispersion $\sigma^2_{x}=\bar{x^2}-\bar{x}^2$ over the filter, to lowest order we have
%--------------
%\beal
%\label{sec:Signal_freq}
%\bar{S}&\approx
%S_{\rm iso}(\bar{x})
%+\frac{1}{2}\partial_{\bar{x}}^2\, S_{\rm iso}(\bar{x})\,\sigma^2_{x}.
%\end{align}
%
$\bar{S}\approx S_{\rm iso}(\bar{x}) +\frac{1}{2}\partial_{\bar{x}}^2\, S_{\rm iso}(\bar{x})\,\sigma^2_{x}$.
%--------------
In Fig.~\ref{fig:S_xx} we show $\partial_{\bar{x}}^2\,S_{\rm iso}(\bar{x})$ for different values of the electron temperature ($\betac=0$).
It is clear that the average of the SZ signal over the filter also leads to another correction term that affects the deduced values for $\tau$, $\TeSZ$, and $\betacpar$.
The significance of the introduced biases depends on the number and position of channels as well as the band-width. At low frequencies the effect is very small, but close to $x\sim 7$ it can reach $\simeq 10\%$ of the main signal for band-width comparable to $10\%$ (assuming a simple top-hat filter). 
\changeJ{As Fig.~\ref{fig:S_xx} shows, the} other two regions with large curvature in the SZ signal are around $x\simeq 3$ and $x\simeq 10-15$, with the latter having a significant dependence on electron temperature.

%---------------
\begin{figure}
\centering
\includegraphics[width=0.94\columnwidth]{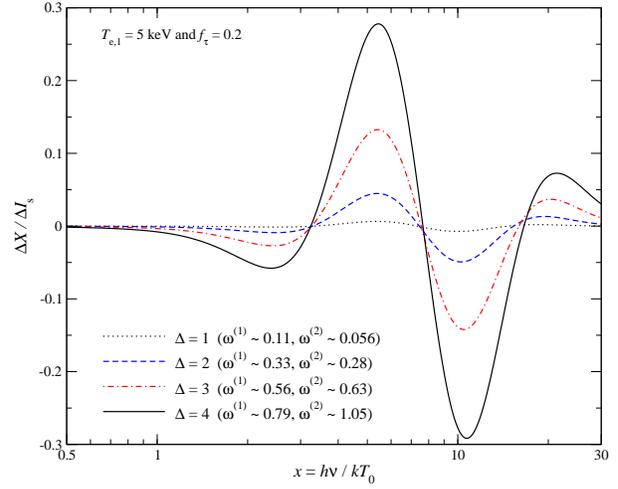}
\caption{Importance of higher order moments for a two-temperature plasma. We assumed $T_{\rm e,1}=5\,\keV$ and $f_\tau=0.2$. The absolute difference of the full two-temperature distortion, $\Delta I_{{\rm two}-T}$, and the approximation $\Delta I_{\rm appr}$ which only uses the average SZ temperature and the temperature dispersion, $\omega^{(1)}$, is shown in units of $\Delta I_{\rm s}\simeq 0.013\, {\rm MJy\,sr^{-1}}$ (i.e., $\Delta X=\Delta I_{{\rm two}-T}-\Delta I_{\rm appr}$).
The temperature difference was varied in each case as annotated.}
\label{fig:two-temp-cases}
\end{figure}
%---------------
\subsection{SZ signal for a two-temperature plasma}
\label{sec:example_two_temp_cluster}
%------------------------------------
In the previous section we illustrated how the SZ signal is affected by the temperature profile of the cluster. However, for the analysis we used very smooth average profiles which allowed us to obtain an accurate description of the SZ signal with only a few new parameters (see Sect.~\ref{sec:example_clusters}).
In general the variance of the temperature, density and velocity field along different lines-of-sight could be much larger than suggested by the simple average profiles.

One illustrative example is to consider a two-temperature plasma, with one low temperature, $T_{\rm e,1}=\Te$, and a high temperature region at $T_{\rm e,2}=\Te(1+\Delta)$, where $\Delta$ is the relative difference between the two temperatures.
We furthermore have to specify the optical depths of each region, $\tau_1=\tau(1-f_\tau)$ and $\tau_2=f_\tau \tau$, where $\tau$ is the total optical depth along the line-of-sight.
From this it follows
%--------------
\beal
\label{sec:two_temp_parameters}
\TeSZ&=\Te (1+f_\tau \Delta)
\nonumber\\
\omega^{(1)}&=f_\tau\frac{(1-f_\tau)\Delta^2}{(1+f_\tau \Delta)^2} < f^{-1}_\tau-1
\\ \nonumber
\omega^{(2)}&=f_\tau\frac{(1-f_\tau)(1-2f_\tau)\Delta^3}{(1+f_\tau \Delta)^3}<(f^{-1}_\tau-1)(f^{-1}_\tau-2)
\end{align}
%--------------
The temperature dispersion has a maximum at $f_\tau=[2+\Delta]^{-1}$ with $\omega^{(1)}_{\rm max}=\frac{1}{4}\Delta^2/[1+\Delta]$. Furthermore, for $\Delta>[1-3f_\tau]^{-1}$ and $f_\tau<1/3$ one finds $\omega^{(1)}<\omega^{(2)}$.

%---------------
\begin{figure*}
\centering
\includegraphics[width=1.5\columnwidth]{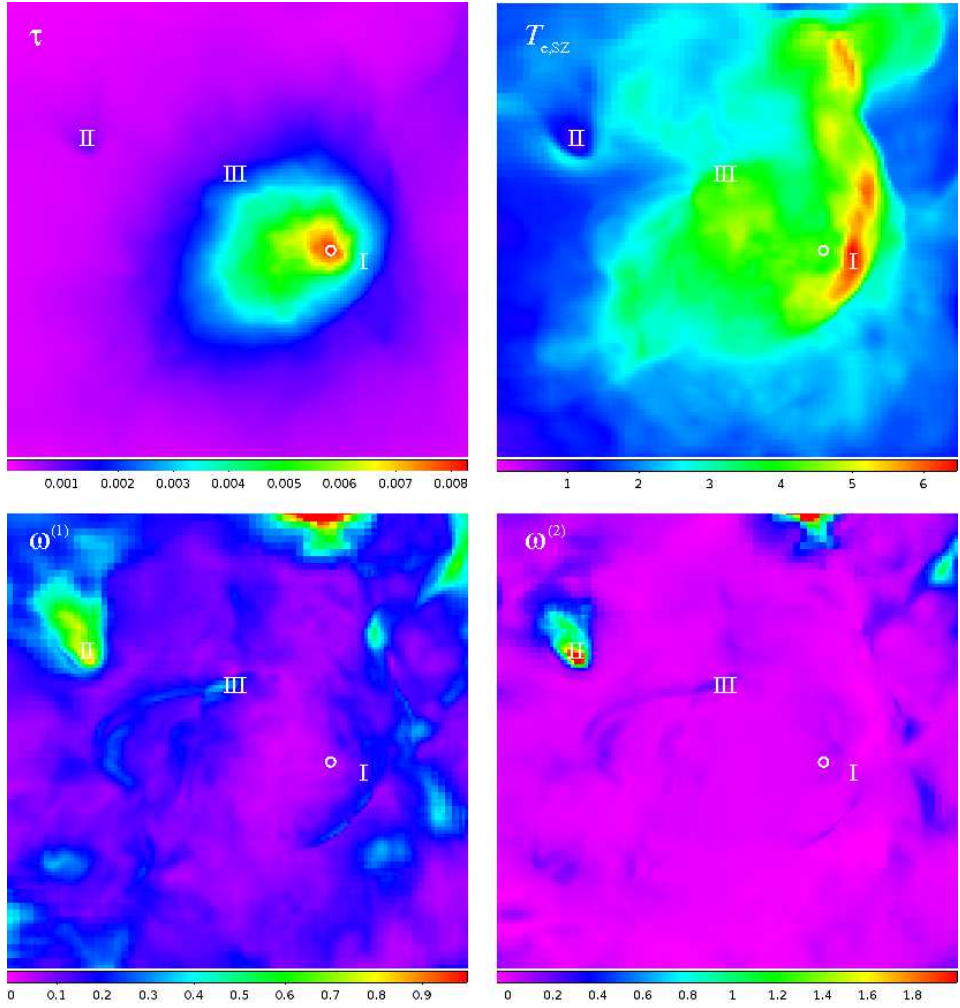}
\caption{First few moments of the temperature field for a simulated cluster from \citet{Nagai2007a, Nagai2007b}. We only show one projection; $\TeSZ$ is in keV. We marked several regions for which we also computed the average SZ signal (see discussion in main text). The highest density peak (marked with an open circle) defines the cluster center. The box is $1\,\Mpc\ h^{-1}$ on the side and the cluster mass is $\simeq 3.8\times 10^{14}\,M_\odot$.}
\label{fig:thSZ_sim}
\end{figure*}
%---------------

In Fig.~\ref{fig:two-temp-cases} we illustrate the importance of higher order temperature moments for a two-temperature plasma with $T_{\rm e,1}$ and $f_\tau=0.2$ and different values of $\Delta$.
For rather small temperature ratio the average SZ signal is approximated very well using the average SZ temperature and $\omega^{(1)}$, with higher order temperature moments leading to very small corrections.
For the most extreme case shown we have $T_{\rm e,1}=5\,\keV$ and $\tau_1=0.8$ with a high temperature component $T_{\rm e,1}=25\,\keV$ and $\tau_2=0.2$. In this case higher order moments are important at the level of a few percent for the SZ signal, but we found that the approximation improved dramatically when only adding $\omega^{(2)}$.
In this example one also had $\omega^{(1)}<\omega^{(2)}$, a case that was not found for the smooth cluster profiles discussed in Sect.~\ref{sec:example_clusters}. Encountering $\omega^{(1)}<\omega^{(2)}$ therefore can be valued as indication of non-smooth, high variance plasma, possibly with more than one temperature component. However, to assess the statistical expectation for such cases detailed simulations of clusters within the cosmological context should be performed.
%

%---------------
\begin{table}
\caption{Optical depth, SZ temperature, and first two line-of-sight temperature moments for the regions marked in Fig.~\ref{fig:thSZ_sim}.}
\label{tab:linesofsight}
\centering
\begin{tabular}{@{}ccccc}
\hline
\hline
Region  & $\tau\,[10^{-3}]$ 
& $\TeSZ \,[\rm keV]$ 
& $\omega^{(1)}$ 
& $\omega^{(2)}$ 
\\
\hline
o & $7.7563$ & $4.1051$ & $0.066421$ & $0.021254$ \\
I & $3.1682$ & $6.3163$ & $0.14538$ & $-0.021378$ \\
II & $0.62251$ & $1.1251$ & $0.69288$ & $1.1797$ \\
III & $1.3738$ & $3.8462$ & $0.29556$ & $0.17844$ \\
\hline 
\hline
\end{tabular}
\end{table}

\subsection{Thermal SZ signal for a simulated cluster}
\label{sec:cluster_sims}
%------------------------------------
We close our discussion of approximations of the SZ signal with one example taken from hydrodynamic simulations presented previously in \citet{Nagai2007a,Nagai2007b}. 
The simulation is performed with the Adaptive Refinement Tree (ART) $N$-body$+$gas-dynamics code \citep{Kravtsov2002,Rudd2008}, which is an Eulerian code that uses adaptive refinement in space and time, and non-adaptive refinement in mass to achieve the dynamic range necessary to resolve the cores of halos formed in self-consistent cosmological simulations. The simulation assumes a flat {$\Lambda$}CDM model: $\Omega_{\rm m}=1-\Omega_{\Lambda}=0.3$, $\Omega_{\rm b}=0.04286$, $h=0.7$ and $\sigma_8=0.9$, where the Hubble constant is defined as $100h{\ \rm km\ s^{-1}\ Mpc^{-1}}$, and $\sigma_8$ is the mass variance within spheres of radius $8\,h^{-1}$~Mpc. The simulation shown here is of a cluster with mass $M_{500}\simeq 3.8\times 10^{14}\ {\rm M_{\odot}}$ undergoing a near 1:1 merger at $z = 0.25$ \citep{Nelson2011}. The simulation was run using non-radiative gas dynamics on a uniform $128^3$ grid with 8 levels of mesh refinement, corresponding to peak spatial resolution of $\simeq 7\,h^{-1}$~kpc.  The dark matter (DM) particle mass in the region around the cluster was $m_{\rm p} \simeq 9.1\times 10^{8}\,h^{-1}\, {\rm M_{\odot}}$.

In Figure~\ref{fig:thSZ_sim} we show projections of $\tau$, $\TeSZ$, $\omega^{(1)}$, and $\omega^{(2)}$. All variables are rather smooth close to the cluster center, but show significant variations in the outskirts. 
We marked regions with temperature variance $\omega^{(1)}\simeq 0.1-0.2$ (see Table~\ref{tab:linesofsight} for details). In these regions also $\omega^{(2)}$ acquired significant values, so that one could expect large contributions to the average SZ signal from higher order terms.
The highest temperature region (marked with `I') is slightly displaced from the maximum of the line-of-sight optical depth (marked by an open circle). It is also coinciding with a region of enhanced temperature variance. 
Region `II' on the other hand is one of the coolest regions in the cluster but very large line-of-sight temperature variation. 

We computed the SZ signal for the different regions using direct integration of the distortion for each frequency.  
We also confirmed that with the moment method developed in Sect.~\ref{sec:Moments} gives the same result but at a significantly lower computational cost.
In both cases, bulk velocity terms were omitted.
We \changeJ{then compared} the approximate SZ signal obtained with the moments summarized in Table~\ref{tab:linesofsight} and found that the deviations were very small, consistent with higher order temperature moments being negligible. This indicates that even for more realistic cluster models the SZ signal is well approximated using the smooth profile expansion, Eq.~\eqref{sec:Signal_eq_tot_series_velocity}. However, the considered cluster generally has rather low temperature so that relativistic corrections are not expected to be as important. We will investigate more massive clusters in our future work.

\section{Effect of different line-of-sight variations on the SZ null}
\label{sec:SZ_null}
\begin{figure}
\centering
\includegraphics[width=0.93\columnwidth]{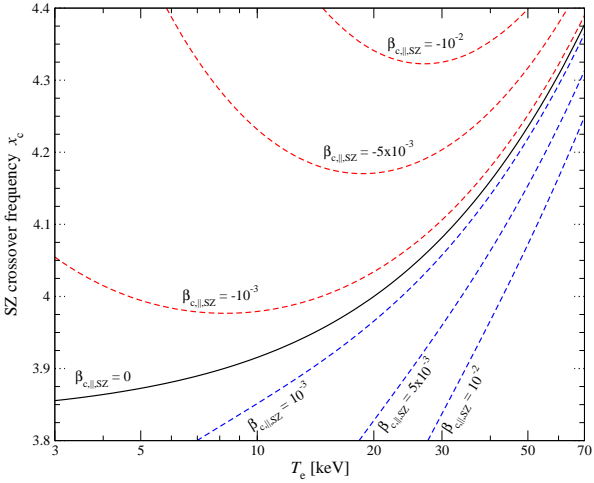}
\caption{Dependence of the crossover frequency on $T_{\rm e, SZ}$ and $\beta_{\rm c, \parallel, SZ}$.}
\label{fig:SZ_null}
\end{figure}

One of the important bands for SZ observations is defined by the SZ null. It is well-known that the position of the \changeJ{so called} crossover frequency depends on the temperature of the cluster gas and average line-of-sight velocity. 
In particular, higher order temperature corrections affect its precise location \citep[e.g., see][]{Nozawa1998}.
This is illustrated in Fig.~\ref{fig:SZ_null}, where the position of the null was computed using one of the functions implemented in {\sc SZpack}. The results are in good agreement with the simple fitting formula given by \citet{Nozawa1998}.

\begin{figure}
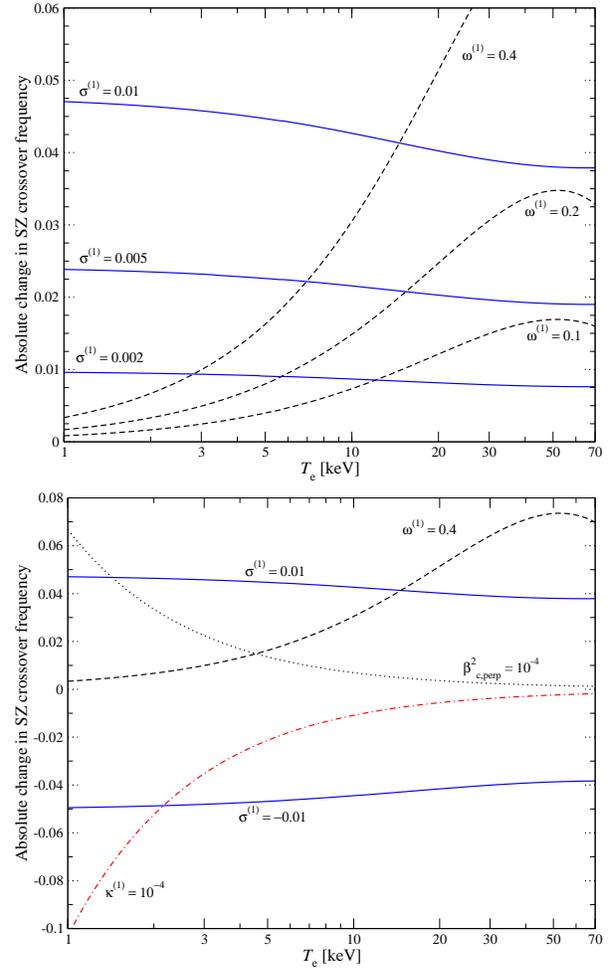

\centering
\includegraphics[width=0.93\columnwidth]{./eps/SZ_null.omega.eps}
\\[1mm]
\includegraphics[width=0.93\columnwidth]{./eps/SZ_null.beta2.eps}
\caption{Dependence of the crossover frequency on $T_{\rm e, SZ}$, $\omega^{(1)}$, $\sigma^{(1)}$, $\kappa^{(1)}$, and $\beta^2_{\rm c,\perp,SZ}$. We present the absolute shift, $x_{\rm c}-x_{\rm c, 0}$, where $x_{\rm c, 0}$ is the case with $\betacpara=\omega^{(1)}=\sigma^{(1)}=\kappa^{(1)}=\beta^2_{\rm c,\perp,SZ}=0$.}
\label{fig:SZ_null.omega}
\label{fig:SZ_null.beta2}
\end{figure}

However, it is clear that the line-of-sight temperature and velocity variations also affect the position of the null (see Fig.~\ref{fig:SZ_null.omega}).  
%
%---------------
%---------------
%---------------
%---------------
Generally, the change in the position of the SZ null is rather small in the cases shown.
The shift is nearly temperature-independent for $\sigma^{(1)}$, while for $\omega^{(1)}$ it grows with $\Te$.
We also present the effect of second order velocity terms on the crossover frequency.
The line-of-sight variance of the velocity causes a decrease of the crossover frequency, while $\beta^2_{\rm c,\perp,SZ}$ leads to an increase.
Both effects are rather small for a typical cluster temperature, even for the relatively large fiducial values that were chosen in the illustration.
They furthermore practically cancel each other, indicating that for SZ parameter estimation they are degenerate.
In the simulated cluster shown in Fig.~\ref{fig:thSZ_sim} we found maximal values of $\beta^2_{\rm c,\perp,SZ}\simeq 10^{-5}$ and $\kappa^{(1)}\simeq 10^{-5}$. However, this depends strongly on the chosen projection and post-merger time. It is therefore important to analyze a larger number of simulated clusters to estimate the typical values for $\beta^2_{\rm c,\perp,SZ}$ and $\kappa^{(1)}$.

\subsection{Effect on the deduced SZ temperature}
The results of Fig.~\ref{fig:SZ_null.omega} indicate that conclusions on the temperature of the cluster drawn from measurements close to the SZ null can be affected by line-of-sight temperature and velocity variations. For instance, assuming $\betacpar=0$, we find 
%--------------
\beal
\xc&\approx 3.83[1+0.011 \Teh-\pot{8.6}{-5} \Teh^2]
\nonumber\\
\Delta \xc &\approx 0.042 \Teh[1 - 0.069 \Teh]\,\omega^{(1)}
\end{align}
%--------------
with $\Teh=\Te/5\,\keV$ for $\Te \lesssim 30\,\keV$.
This means that for SZ temperature $\TeSZ=5\,\keV$ and $\omega^{(1)}\simeq 0.2$, we find $\xc\approx 3.88$, which would imply a temperature of $\TeSZ \simeq 6\,\keV$ when assuming $\omega^{(1)}=0$ in the analysis. This is a $\simeq 19\%$ bias towards higher temperature.
More generally we find a bias of
%--------------
\beal
\label{eq:bias_Te_xc_omega}
\frac{\Delta \Te}{\Te}\approx \frac{1-0.069 \Teh}{1-0.016 \Teh} \,\omega^{(1)}
\end{align}
%--------------
for $\Te \lesssim 30\,\keV$. The degeneracy between $\Te$ and $\omega^{(1)}$ is very close to unity at low temperatures; however, for $\TeSZ=30\,\keV$ and $\omega^{(1)}\simeq 0.2$ the temperature bias reduces to $\simeq  13\%$, which is also in good agreement with the behaviour found in Fig.~\ref{fig:degen_coeffy}.
Although the absolute shift in $\xc$ caused by $\omega^{(1)}$ increases with temperature (cf. Fig.~\ref{fig:SZ_null.omega}), the relative shift decreases.

One can also estimate the effect of higher order velocity terms on the inferred value of $\TeSZ$. For contributions related to $\mathcal{O}(\betac^2)$ the effect is very small and only relevant for very low temperature gas, if at all. 
Similarly, the value of $\sigma^{(1)}$ is expected to be much smaller than $\lesssim 10^{-3}$, unless high temperature gas also has large bulk velocity.
For example, for the simulated cluster shown in Fig.~\ref{fig:thSZ_sim} we found maximal values of $\sigma^{(1)}\simeq \pot{4.7}{-4}$.
Therefore the correction caused by $\omega^{(1)}$ is expected to be dominant, unless the overall temperature of the electron gas is very low.

As explained in Sect.~\ref{sec:angular_res}, also variations caused by averaging the SZ signal over the beam introduce a dispersion. This has the same effect as the line-of-sight temperature variation, but for small beam the latter should again dominate.

\subsection{Effect of frequency resolution on the SZ null}
\label{sec:frequencies_null}
%--------------------------------
It is also straightforward to estimate the effect of the bandpass on the SZ null. The shift in its position is approximately given by
%--------------
\beal
\label{sec:xc_shift_freq}
\Delta \xc
&\approx - \frac{\partial_{\xc}^2 S(\xc)}{2\,\partial_{\xc} S(\xc)}\,\sigma^2_{x} 
\nonumber\\
&\approx -0.2 [1 -0.32 \Teh+0.031\Teh^2-0.0018\Teh^3]\,\sigma^2_{x}
\end{align}
%--------------
for $\Te\lesssim 30\,\keV$. At $\Te\simeq 23\,\keV$ the shift in the crossover frequency caused by the average over the filter becomes small. Also, for a top-hat frequency filter one has $\sigma^2_{x}= \Delta x^2 /12\simeq 1.22 (\Delta \nu/\nu_{\rm c})^2$, which for a bandwidth of $10\%$ and $\Te \simeq 5\,\keV$ implies a shift of $\Delta \xc \simeq -0.002$ in the crossover frequency. This is expected to result in a $\simeq 4\%$ bias of $\Te$ towards lower values.
More generally we find
%--------------
\beal
\label{eq:bias_Te_xc_f}
\frac{\Delta \Te}{\Te}\approx -\frac{5.8}{\Teh}
\left(\frac{\Delta \nu}{\nu_{\rm c}}\right)^2
\frac{1-0.298 \Teh+0.024 \Teh^2-0.0011\Teh^3}{1-0.016 \Teh} 
\end{align}
%--------------
caused by the band-width of the frequency channel around $\xc(\Te)$. 

One should mention that we assumed that the frequency band is centered exactly on the crossover frequency, which of course changes with temperature, $\Te$. 
If the central frequency differs from $\xc$ the response is also affected and hence the deduced temperature $\Te$. In the limit of very narrow bandpass this problem disappears, but otherwise this effect should be taken into account when interpreting SZ data at high precision.

\section{SZ parameter estimation}
\label{sec:SZ_parameter_est}
%------------------------------
In Sec.~\ref{sec:Moments}, we developed a new method for accurately calculating the mean SZ signal with relativistic corrections, and in Sec.~\ref{sec:SZclusters} we extended this to include fluctuations in temperature and velocity along the line-of-sight for smooth cluster profiles. Subsets of the model parameters $\{ \tau, \TeSZ, \omega^{(1,2,3)}, \sigma^{(1,2,3)}, \kappa^{(1,2,3)}, \betacparaSZ, \betacperpSZ^2 \}$, can then be estimated from measurements of the SZ spectrum\footnote{We will drop the subscript `SZ' for convenience, but it is important to bear in mind that for instance the derived electron temperature, $\Te\equiv\TeSZ$, i.e. it is weighted by the electron number density.}. 
Here we describe an estimation of these parameters and their interpretation using a compilation of data and simulations. The parameters are estimated in a Markov Chain Monte Carlo \citep{Foreman2012} which is available with {\sc SZpack} and can be adapted for particular experimental settings.

Rather than performing an exhaustive survey of experimental configurations and considerations, we describe a few informative cases. Even for a given combination of multi-band data, a complete interpretation is beyond the scope of this paper because it demands a consideration of contaminants, calibration errors, and offsets which can all be correlated across the bands. Particulars of the observation strategy may also need to be fully modelled \citep[e.g., see][]{Zemcov2012}. The examples here are therefore meant to show some degeneracies and interpretative issues that arise in multi-band SZ parameter estimation. The same conclusions broadly apply to stacked cluster studies, except there, for example, the kSZ may average to zero.

\begin{figure}
\includegraphics[width=0.96\columnwidth]{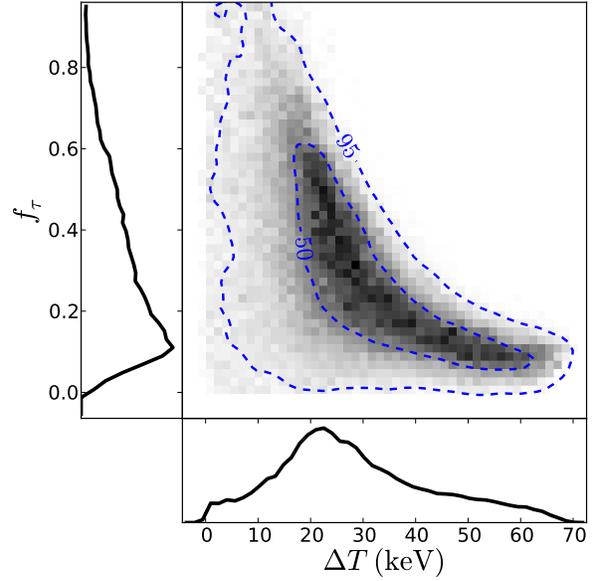}
\caption{Joint marginalized posterior distribution for a two-temperature model of the Bullet cluster showing the fractional optical depth $f_\tau$ which can occupy a component of gas with temperature $\Delta T$ higher than the main component. Here, the temperature of the main component is $(7.2 \pm 3.9)\,\keV$. When the additional hot component is $20$~keV hotter, it can occupy a significant fraction of the optical depth while for larger excursions in temperature is must have progressively smaller optical depth. Note that non-zero $\Delta T$ is driven by the measurement at $857$~GHz, where point source contamination could be significant.\label{fig:twotemp}}
\end{figure}

\subsection{Interpreting temperature variance in a recent SZ data compilation for the Bullet cluster}
\label{sec:Prokhorov}
%------------------------
The simplest estimation problem which demonstrates evidence of temperature dispersion is the two-temperature model of Sec.~\ref{sec:example_two_temp_cluster}. Here the parameters are a reference temperature $\Te$, a temperature difference $\Delta T$, and the fractional optical depth \changeJ{$f_\tau<1$} in the region with temperature $\Te + \Delta T$. 
A two-temperature case was also recently considered by \citet{Colafrancesco2011} as model for the Bullet cluster gas.
The two SZ contributions from the gas at $\Te $ and $\Te + \Delta T$ are computed to all relevant relativistic correction orders using {\sc SZpack}. Following the compilation of \citet{Prokhorov2012}, we consider data across the bands $\{150, 275, 600, 857\}\,{\rm GHz}$ with mean values and errors of $\changeJ{\Delta I=}\{-0.325 \pm 0.015, 0.21 \pm 0.077, 0.268 \pm 0.031, 0.097 \pm 0.019\}\,\rm MJy/sr$ from observations of \citet{Gomez2004} and \citet{Zemcov2010}. We assume that $\tau = 0.0138 \pm 0.0016$ is known \citep{Prokhorov2012} as a Gaussian prior probability distribution from X-ray observations. This helps break the strong degeneracy between $\tau$ and $\Te$. 
For the latter we find $\Te=(7.2 \pm 3.9)\,\keV$.
Figure~\ref{fig:twotemp} shows the marginalized posterior distribution for $\Delta T = 25^{+19}_{-11}$~keV and $f_\tau = 0.27^{+0.26}_{-0.17}$ (where two-sided errors are given for non-Gaussian distributions at percentiles $\{ 16 \%, 50 \%, 84\% \}$, equivalent to $1\sigma$ in a normal distribution). Conclusions from a model with $\betacpara$ (taken to be the same for both components) as a free parameter are similar at this level of uncertainty.

\begin{figure}
\includegraphics[width=0.96\columnwidth]{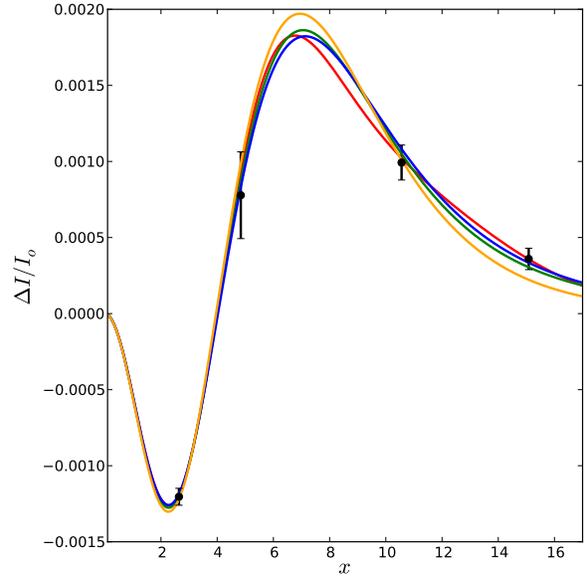}
\caption{Mean SZ spectra for four models of the data (black points) \citep{Prokhorov2012}. A model including only $\Te$ with relativistic corrections at all relevant orders (yellow) does not have the freedom to describe the $857$~GHz flux, and drives the temperature up. An asymptotic expansion of relativistic corrections to order $\langle (k \Te)^3 \rangle/(\me c^2)^3$ (red) describes the data, but is insufficiently converged to self-consistently explain the relativistic corrections to the SZ flux. The green curve is a model with two regions of different temperature and the blue curve is a model with dispersion about a mean temperature (both cases include all relevant orders in the relativistic corrections).  Measurements around and above the SZ maximum have the potential to differentiate these models, while close to the SZ null and below all models practically coincide.
\label{fig:mean_models}}
\end{figure}

We next consider an ICM model with line-of-sight temperature dispersion about the mean using the free parameters $\{\tau, \Te, \omega^{(1)} \}$, where $\omega^{(1)} = \langle \Te^2 \rangle / [\tau \TeSZ^2] - 1$ is a measure of the temperature dispersion \changeJ{(see Sect.~\ref{sec:example_clusters})}. In this case, we obtain $\Te = (13.5 \pm 1.8)\,\keV$ and $\omega^{(1)} = 0.8 \pm 0.4$ with a modest correlation ${\rm corr}(\Te, \omega^{(1)}) = -0.4$. 
When $\betacpara$ is added as a free parameter, we find $\Te = (13.1 \pm 2.0)\,\keV$ and $\omega^{(1)} = 1.0 \pm 0.6$ with $\betacpara = (-1.7 \pm 2.1) \times 10^{-3}$ and ${\rm corr}(\betacpara, \omega^{(1)}) = -0.6$, showing that there is no significant indication of non-zero line-of-sight bulk velocity. 
We can extend the fit to include $\omega^{(2)}$, a measure of the skewness in the temperature distribution. Here, $\omega^{(1)} = 1.1 \pm 0.6$ and $\omega^{(2)} = 2.0 \pm 2.6$ (no strong evidence for more than dispersion) with a strong correlation, ${\rm corr}(\omega^{(1)},\omega^{(2)}) = 0.75$. It is also informative to compare with a model where $\omega^{(1)}$ is fixed to zero and only $\Te$ and $\tau$ vary. In this case, $\Te = (14.8 \pm  2.1)\,\keV$, but as shown in Fig.~\ref{fig:mean_models}, this is a poorer fit to the data because it lacks the freedom to explain the flux measurement at $857$~GHz.

The two previous models depart from the first reported measurement of temperature dispersion by \citet{Prokhorov2012}, where the temperature expansion is used directly to constrain the variance:
%--------------------
\beal
\label{eqn:CLPC}
\frac{\Delta I}{\changeC{I_{\rm o}}} &\approx 
\tau \biggl [ g_0(x) \,\frac{\langle k \Te \rangle}{\me c^2} 
+ g_1(x) \,\frac{\langle (k \Te)^2 \rangle}{\me^2 c^4} 
+ g_2(x)\, \frac{\langle (k \Te)^3 \rangle}{\me^3 c^6} \biggr ]. 
\end{align}
%--------------------
Here $g_0$, $g_1$ and $g_2$ are the spectral functions of the asymptotic expansion given in \citet{Challinor1998} and flux normalization, $\changeC{I_{\rm o}}\equiv (2h/c^2) (kT_0/h)^3\approx 270\,\MJysr$. The temperature variance can be inferred directly through $\sigma^2 = \langle (k \Te)^2 \rangle - \langle k \Te \rangle^2$, by fitting for the coefficients of $g_0$, $g_1$ and $g_2$. As a check of the MCMC here, we find a temperature variance of $(9.2 \pm 2.5)$~keV, in agreement with $(9.5 \pm 2.6)$~keV reported in \citet{Prokhorov2012}. The average temperature of the Bullet cluster inferred from X-ray observations \citep{Million2009} is $14.5$~keV. At these temperatures, Fig.~\ref{fig:precision_Itoh_CNSN} shows that the asymptotic expansion is a poor approximation even to order $\langle (k \Te)^{10} \rangle/(\me c^2)^{10}$, which worsens for higher temperature excursions. Because the first few moments do not adequately describe the mean relativistic corrections, it is also difficult to interpret the dispersion inferred from the first two moments in Eq.~\eqref{eqn:CLPC}. Nevertheless, converting the moment constraints, we find that $\Te = (13.7 \pm 1.9)$~keV and $\omega^{(1)} \sim\sigma^2 / \Te^2 = (0.5 \pm 0.3)$ for this model, which is consistent with the dispersion measured above.
However, as the experimental sensitivity and frequency coverage increases the model given by Eq.~\eqref{eqn:CLPC} should lead to biased values for $\Te$ and $\omega^{(1)}$.
%

%------------------------------
%------------------------------

In any of the models above, the detection of dispersion is driven by the flux measurement at $857$~GHz. Removing that measurement from the two-temperature model parameter estimation results in an upper bound rather than detection ($\Delta T < 56$~keV, $95\%$ confidence). In the dispersion model with parameters $\{ \Te, \tau, \omega^{(1)} \}$, we find $\omega^{(1)} = 0.4 \pm 0.7$.
Most of the constraint on relativistic effects and dispersion comes from above the SZ null, but in this frequency region it becomes more difficult to separate point source contamination, both because IR fluxes increase at higher frequency and because both the SZ and contamination are positive fluxes.

The parameter estimates here are based on flux measurements in four frequencies, and yield qualitatively similar conclusions for the current sensitivity. Across a range of frequencies, however, the three models do predict different SZ spectra, shown in Fig.~\ref{fig:mean_models}. Here, the expression of Eq.~\eqref{eqn:CLPC} shows the most variation with respect to other models with temperature dispersion because three temperature moments in the asymptotic series do not fully describe the relativistic corrections and temperature dispersion. The other two models show slight differences above the SZ maximum, indicating that more precise measurements at a wider range of frequencies could help differentiate between physical interpretations of the dispersion. 
Still the differences in the SZ signal above the null are not much larger than $\simeq 5\%-10\%$.
There are limited transparency windows in the atmosphere, but at frequencies above the null, instruments can achieve similar resolution and sensitivity with comparatively smaller apertures, so that balloon and space-borne instruments could have the potential to constrain these frequency regions. Note that the distortion around the null is virtually identical in all the models described here.

%\subsection{Additional mock observations}
\label{sec:mock}
%%------------------------------
In the constraints above, we neglected the finite bandwidth of the observations. It is customary to define an effective band center for a given spectral index as the frequency at which the measured flux matches the true flux. We can expand smooth spectra about this effective band center and see that the linear term drops out and that the change in center is due to the second derivative curvature of the spectrum across the band (see Sect~\ref{sec:frequencies_null}). The extent of this effect depends on experiment-dependent parameters, but as a rough estimate, a $10\%$ bandwidth near the maximum at $x \sim 7$ can produce a nearly $10 \%$ bias. In the data considered above, for a $10\%$ bandwidth, we find $\Te = (13.6 \pm 1.8)$~keV and $\omega^{(1)} = 0.72 \pm 0.4$, a shift which can be neglected here, but should become important as the measurements improve.

\begin{figure}
\includegraphics[width=\columnwidth]{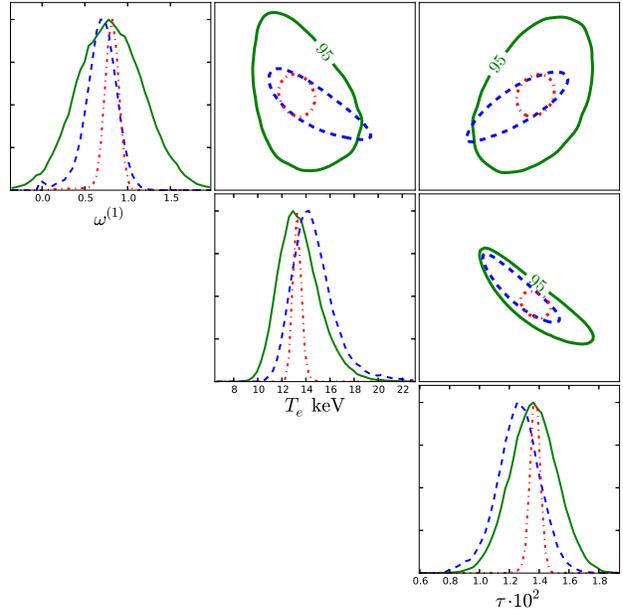}
\caption{Posterior distributions of $\{ \Te, \tau, \omega^{(1)} \}$ achieved by improvements to the measurements compiled in \citet{Prokhorov2012}. The green solid line gives 1D marginalized probabilities and 2D contours ($95\%$ enclosed) for the original dataset. This shows a clear indication of scatter about the central temperature through significantly non-zero $\omega^{(1)}$. Improving the SZ flux measurements by a factor of five (keeping the measured central values) only improves the constraint on $\Te$ and $\omega^{(1)}$ (blue dashed line) modestly because of their degeneracy with $\tau$. Additionally shrinking the $\tau$ prior errors by a factor of five (red dot-dashed line) breaks this to yield the strongest constraints. \label{fig:noise_summary}}
\end{figure}

We can also examine the impact of improved measurements on the inference of $\omega^{(1)}$. Fig.~\ref{fig:noise_summary} considers the parameter set $\{\tau, \Te, \omega^{(1)}\}$ under improvements to the dataset used in \citet{Prokhorov2012}. By reducing the SZ measurement errors a factor of five for the same $\tau$ prior, the $\omega^{(1)}-\Te$ correlation increases to $-0.7$ and the parameter errors improve modestly. Both the correlation and residual uncertainty are driven by the imperfect $\tau$ prior, and deeper SZ measurements have little additional benefit. Shrinking the $\tau$ prior by a factor of five on top of the improved SZ flux measurements significantly improves these constraints.
This also emphasizes the importance of combining SZ measurement with X-ray observations to break degeneracies among model parameters.

\section{Discussion and conclusion}
\label{sec:disc-conc}
%------------------------------
We developed a novel method for extracting information about the state of the ICM from SZ observations which is based on moments of the cluster's temperature and velocity field.
Both a quasi-exact approach (Sect.~\ref{sec:Moments}), which should be useful for computing the precise SZ signal from simulated clusters in an efficient manner, and an approximate treatment based on the smoothness of the \changeJ{cluster profile} (Sect.~\ref{sec:example_clusters}) were considered. 
In future SZ parameter estimation and cluster profile reconstruction the latter method should be widely applicable, with the main SZ observables consisting of a small number of parameters, $\vek{p}=\{ \tau, \TeSZ, \omega^{(1,2,3)}, \sigma^{(1,2,3)}, \kappa^{(1,2,3)}, \betacparaSZ, \betacperpSZ^2 \}$.
Maps of these parameters allow very precise representation of the SZ signal and hence a compression of the observational data to only spatially dependent variables. 
These can be directly used in the inversion problem that determines the cluster temperature and velocity structure.

Our discussion of different cluster models indicated that the above compression of the data should work even in more general cases, when the line-of-sight temperature and velocity distributions become non-trivial. The reason is simply that the SZ signal itself is very smooth in frequency and higher order temperature and velocity moments contribute in a diminishing manner, once the average SZ signal (which includes a large number of temperature correction terms) is taken out.
We also showed how the position of the SZ null is affected by temperature variations along the line-of-sight. If neglected this can lead to biases in the deduced electron temperature at the level of $\simeq 10\%-20\%$ (see Eq.~\ref{eq:bias_Te_xc_omega}).
Velocity correction terms, beyond the normal kSZ effect, are generally expected to be less important; however, a more detailed study using hydrodynamic cluster simulations is needed to give a more quantitative answer about the typical magnitude of different terms.

We furthermore illustrated the importance of high frequency channels for distinguishing different cluster atmospheres (Fig.~\ref{fig:mean_models}) using recent data for the Bullet cluster \citep{Prokhorov2012}. Our analysis indicates that presently at two-temperature model remains indistinguishable from a cluster model with line-of-sight temperature variance, but a single temperature model is in tension with the measured SZ flux at $857\,\GHz$. We also explain why the results for the temperature dispersion deduced with an asymptotic expansion of the SZ signal are difficult to interpret.
In the future, it will be important to consider foregrounds and systematic effects in more detail. Also additional physical effect, e.g., multiple-scattering effects \citep{Dolgov2001} or the SZ effect for non-thermal electron populations \citep{Ensslin2000}, should be taken into account. 
We plan to extend {\sc SZpack} with these features in another publication.

To realize the full power of SZ measurement as a cosmological probe, studies of individual clusters and their atmosphere will become very important. It is especially critical to extend the studies of thermodynamic and velocity structures to the outskirts of high-redshift clusters in order to understand the biases in the global properties of galaxy clusters and their evolution across the cosmic time.
Because the SZ signal is independent of redshift and linearly proportional to the gas density, high-resolution SZ observations can enable detailed studies of the ICM structure in the outskirts of high-redshift clusters, and hence are highly complimentary to X-ray observations whose sensitivities are limited to the inner regions of clusters or the outskirts of nearby systems.
The combination of future spectroscopic X-ray measurements with high resolution SZ data will open new opportunities for studying cluster astrophysics, and the moment method developed here provides one important step in this direction.

\changeJ{\section*{Acknowledgments}
We thanks Eiichiro Komatsu, Andrea Morandi, and Michael Zemcov for valuable comments and discussion. DN and KN acknowledge the support from the NSF grant AST-1009811, the NASA  ATP grant NNX11AE07G, the NASA Chandra Theory grant GO213004B, Research Corporation, and by Yale University. This work was supported in part by the facilities and staff of the Yale University Faculty of Arts and Sciences High Performance Computing Center. 
}

\begin{appendix}

\section{Definition of functions for asymptotic expansion}
\label{app:der_asym}
%-------------------------------
From Eq. (25) of CNSN we can directly identify
%--------------
\beal
\label{eq:SZ_signal_functions_defs}
M^{\rm low}_k &= Y^{\rm kin}_k +\frac{1}{6} \mathcal{D}_{x} Y_k
 -\frac{1}{3}\mathcal{D}^\ast_{x} D^{\rm kin}_k 
\nonumber\\
D^{\rm low}_k&=D^{\rm kin}_k - x \, \partial_x Y_k 
\\\nonumber
Q^{\rm low}_k&=Q^{\rm kin}_k +\frac{1}{3} x^2 \partial^2_x Y_k
 +\frac{2}{3}\mathcal{D}^{\ast\ast}_{x} D^{\rm kin}_k.
\end{align}
%--------------
The functions $Y_k$, $Y^{\rm kin}_k$, $D^{\rm kin}_k$, and $Q^{\rm kin}_k$ are all defined in CNSN.
The differential operators are $\mathcal{D}_{x}=3 x\, \partial_x + x^2 \partial^2_x$, $\mathcal{D}^\ast_{x}=2+ x\, \partial_x$, and $\mathcal{D}^{\ast\ast}_{x}=1-x\, \partial_x$.
With the relations given in CNSN it is straightforward to show that
%--------------
\beal
\label{eq:SZ_signal_functions_derivs}
x\,\partial_x Y_n&=
\sum_{k=1}^{2n+2} a_k^{(n)} \left[k\,\Delta^k_{\xg}+\Delta^{k+1}_{\xg}\right] \nPl(x)
\nonumber\\
\nonumber
x^2\partial^2_x Y_n&=
\sum_{k=1}^{2n+2} a_k^{(n)} \left[k(k-1)\,\Delta^k_{\xg}+2k\,\Delta^{k+1}_{\xg}+\Delta^{k+2}_{\xg}\right] \nPl(x)
\\
\mathcal{D}_x Y_n&=6 \,Y^{\rm kin}_n
\\\nonumber
\mathcal{D}^\ast_x D^{\rm kin}_n&=
\sum_{k=0}^{2n+2} d_k^{(n)} \left[k(k+2)\,\Delta_x^k+(2k+3)\,\Delta_x^{k+1}+\Delta_x^{k+2}\right] \nPl(x)
\\\nonumber
\mathcal{D}^{\ast\ast}_x D^{\rm kin}_n&=-
\sum_{k=0}^{2n+2} d_k^{(n)} \left[k(k-1)\,\Delta^k_{\xg}+2k\,\Delta^{k+1}_{\xg}+\Delta^{k+2}_{\xg}\right] \nPl(x)
\end{align}
%--------------
with $\nPl(x)=1/[e^x-1]$ and $\Delta_x^k=\xg^{k}\partial^k_{\xg}$. This implies
%--------------
\beal
\label{eq:SZ_signal_functions_defs_explicit}
M^{\rm low}_n &= 
\frac{1}{3}\sum_{k=0}^{2n+2} (a_k^{(n)}-d_k^{(n)})
\left[k(k+2)\,\Delta^k_{\xg}+(2k+3)\,\Delta^{k+1}_{\xg}+\Delta^{k+2}_{\xg}\right] \nPl
\nonumber\\
D^{\rm low}_n&=
\sum_{k=0}^{2n+2} (d_k^{(n)}-a_k^{(n)}) 
\left[k\,\Delta^k_{\xg}+\Delta^{k+1}_{\xg}\right] \nPl(x)
\\\nonumber
Q^{\rm low}_n&=
\frac{1}{3}\sum_{k=0}^{2n+2} (a_k^{(n)}+q_k^{(n)}-2d_k^{(n)})
\left[k(k-1)\,\Delta^k_{\xg}+2k\,\Delta^{k+1}_{\xg}+\Delta^{k+2}_{\xg}\right] \nPl
\end{align}
%--------------
We extended the sums for $Y_k$ and its derivatives to $k=0$ using $a_0^{(n)}=0$ for $n\geq 0$.

\section{Definition of functions for the improved basis set}
\label{app:der_new}
%-------------------------------
It is straightforward to transform the set of basis functions given in CNSN from the cluster frame into the CMB rest frame.
However, to obtain the expressions for $Y^{\rm high}_k$, $M^{\rm high}_k$, $D^{\rm high}_k$, and $Q^{\rm high}_k$, a few intermediate steps are needed.
Using the definitions of CNSN \changeJ{gives}
%--------------
\beal
\label{eq:SZ_signal_functions_defs_new}
Y^{\rm high, \ast}_k &= Z_k
\nonumber\\
M^{\rm high, \ast}_k &= Z^{\rm kin}_k +\frac{1}{6} \mathcal{D}_{x} Z_k
 -\frac{1}{3}\mathcal{D}^\ast_{x} C^{\rm kin}_k 
\nonumber\\
D^{\rm high, \ast}_k&=C^{\rm kin}_k - x \, \partial_x Z_k 
\\\nonumber
Q^{\rm high, \ast}_k&=S^{\rm kin}_k +\frac{1}{3} x^2 \partial^2_x Z_k
 +\frac{2}{3}\mathcal{D}^{\ast\ast}_{x} C^{\rm kin}_k.
\end{align}
%--------------
These can be used like the functions $Y^{\rm high}_k$, $M^{\rm high}_k$, $D^{\rm high}_k$, and $Q^{\rm high}_k$ in Eq.~\eqref{eq:SZ_signal_functions_hot} but with $\mathcal{N}(\The)\, \The^k\rightarrow \mathcal{N}(\The)\, (\The-\theta_{\rm e,0})^k /\mathcal{N}(\theta_{\rm e,0})$.
Since
%--------------
\beal
\label{eq:zk_variable_transform}
\frac{\mathcal{N}(\The)}{\mathcal{N}(\theta_{\rm e, 0})} \left(\The-\theta_{\rm e, 0}\right)^{k}\!  \equiv \frac{1}{\mathcal{N}(\theta_{\rm e, 0})}\,\sum_{m=0}^k 
\binom{k}{m} \, (-\theta_{\rm e, 0})^{k-m}\,\mathcal{N}(\The)\,\The^m
\end{align}
%--------------
by defining the lower triangular matrix, $T_{ij}=\frac{1}{\mathcal{N}(\theta_{\rm e, 0})}\, 
\binom{i}{j} \, (-\theta_{\rm e, 0})^{i-j}$, we have the transformation $\vek{X}^{\rm high} = \matrixf{T}^T \vek{X}^{\rm high, \ast}$ to the basis $Y^{\rm high}_k$, $M^{\rm high}_k$, $D^{\rm high}_k$, and $Q^{\rm high}_k$.
Here we used $(\vek{X}^{\rm high, \ast})^T=(X^{\rm high, \ast}_0, ..., X^{\rm high, \ast}_{\rm kmax})$ and $(\vek{X}^{\rm high})^T=(X^{\rm high}_0, ..., X^{\rm high}_{\rm kmax})$, with $X\in\{Y, M, D, Q\}$.

Finally, with the definitions of CNSN, the integrals and derivatives in Eq.~\eqref{eq:SZ_signal_functions_defs_new} can be written as
%-----------
\beal
\label{eq:SZ_signal_functions_defs_new_explicit}
%I_{lm}^{st}= \frac{1}{\Ne\,\sigT}\int \frac{\id\sigma}{\id \Omega'}
%\, f(\vp)\,\frac{\Delta^{s+t}}{s!t!}\,Y_{lm}(\vghp) \,\id^2 \hat{\gamma}' \id^3 p 
M^{\rm high}_k(x)&=   \frac{\mathcal{N}}{3 k!}\!\!
\int\!\! \frac{\id^2 \sigma_{0}}{\id\mu \id \mu'}
\frac{\partial^k e^{-\Delta\gamma/\theta_{\rm e}^{\rm c}}}{\partial^k \The^{\rm c}}
\left[\mathcal{M}(x')-\mathcal{M}(x)\right]\id\mu \id\mu' \eta ^2\!\id \eta, 
\nonumber
\\
\nonumber
&\qquad-\frac{\mathcal{N}}{3k!}\!\!
\int\!\! \frac{\id^2\sigma_{1}}{\id\mu \id \mu'}
\frac{\partial^k e^{-\Delta\gamma/\theta_{\rm e}^{\rm c}}}{\partial^k \The^{\rm c}}
\mathcal{M}(x') \id\mu \id\mu' \eta ^2\!\id \eta, 
\\[1mm]
D^{\rm high}_k(x)&= 
-\frac{\mathcal{N}}{k!}\!\!
\int\!\! \frac{\id^2\sigma_{1}}{\id\mu \id \mu'}
\frac{\partial^k e^{-\Delta\gamma/\theta_{\rm e}^{\rm c}}}{\partial^k \The^{\rm c}}
\,\mathcal{G}(x')\id\mu \id\mu' \eta ^2\!\id \eta
\\
\nonumber
&\qquad+\frac{\mathcal{N}}{k!}\!\!
\int\!\! \frac{\id^2\sigma_{0}}{\id\mu \id \mu'}
\frac{\partial^k e^{-\Delta\gamma/\theta_{\rm e}^{\rm c}}}{\partial^k \The^{\rm c}}
[\mathcal{G}(x')-\mathcal{G}(x)] \id\mu \id\mu' \eta ^2\!\id \eta, 
\\[1mm]
\nonumber
S^{\rm kin}_k(x)&=   
\frac{\mathcal{N}}{3k!}\!\!
\int\!\! \frac{\id^2 \sigma_{2}}{\id\mu \id \mu'}
\frac{\partial^k e^{-\Delta\gamma/\theta_{\rm e}^{\rm c}}}{\partial^k \The^{\rm c}}
\,[\mathcal{Q}(x') - \Lambda_0\mathcal{Q}(x)]
\id\mu \id\mu' \eta ^2\!\id \eta
\\
\nonumber
&\quad+
\frac{\mathcal{N}}{3k!}\!\!
\int\!\! \frac{\id^2 \sigma_{0}}{\id\mu \id \mu'}
\frac{\partial^k e^{-\Delta\gamma/\theta_{\rm e}^{\rm c}}}{\partial^k \The^{\rm c}}
\,[\mathcal{Q}(x') -\mathcal{Q}(x) ]
\id\mu \id\mu' \eta ^2\!\id \eta
\\
\nonumber
&\qquad-
\frac{2\mathcal{N}}{3k!}\!\!
\int\!\! \frac{\id^2\sigma_1}{\id\mu \id \mu'}
\frac{\partial^k e^{-\Delta\gamma/\theta_{\rm e}^{\rm c}}}{\partial^k \The^{\rm c}}
\,\mathcal{Q}(x')
\id\mu \id\mu' \eta ^2\!\id \eta.
\end{align}
%-----------
We implemented these alternative basis functions for {\sc SZpack}.

\end{appendix}

\footnotesize
\bibliographystyle{mn2e}
\bibliography{Lit}
%---------------
\end{document}

%% file: Befehle.tex
\newcommand{\vbetac}{{{\boldsymbol\beta}_{\rm c}}}
\newcommand{\vbetach}{{\hat{{\boldsymbol\beta}}_{\rm c}}}
\newcommand{\betac}{\beta_{\rm c}}
\newcommand{\betacpar}{{\beta_{\rm c,\parallel}}}

\newcommand{\betacsq}{{\beta^2_{\rm c}}}

\newcommand{\muc}{\mu_{\rm c}}

\newcommand{\vgh}{{\hat{\boldsymbol\gamma}}}

\newcommand{\xg}{x}

\newcommand{\xc}{x_{\rm c}}

\newcommand{\id}{{\,\rm d}}

\newcommand{\beq}{\begin{equation}}   %

\newcommand{\eeq}{\end{equation}}   %

\newcommand{\beqa}{\begin{eqnarray}}   %

\newcommand{\eeqa}{\end{eqnarray}}   %

\newcommand{\beal}{\begin{align}}

\newcommand{\enal}{\end{align}}

\newcommand{\bspl}{\begin{split}}

\newcommand{\espl}{\end{split}}

\newcommand{\bsub}{\begin{subequations}}

\newcommand{\esub}{\end{subequations}}

\newcommand{\bmulti}{\begin{multline}}   %

\newcommand{\beqm}{\begin{mathletters}}   %

\newcommand{\eeqm}{\end{mathletters}}   %

\newcommand{\me}{m_{\rm e}}

\newcommand{\Ne}{N_{\rm e}}

\newcommand{\Te}{T_{\rm e}}

\newcommand{\The}{\theta_{\rm e}}

\newcommand{\nPl}{n_{\rm Pl}}

\newcommand{\vek} [1]{\mbox{\boldmath${#1}$\unboldmath}}

\newcommand{\pot}[2]{#1 \times 10^{#2}}

%% file: paper.bbl
\begin{thebibliography}{70}
\expandafter\ifx\csname natexlab\endcsname\relax\def\natexlab#1{#1}\fi

\bibitem[{{Benson} {et~al}\mbox{.}(2004){Benson}, {Church}, {Ade}, {Bock},
  {Ganga}, {Henson}, \& {Thompson}}]{Benson2004}
{Benson} B.~A., {Church} S.~E., {Ade} P.~A.~R., {Bock} J.~J., {Ganga} K.~M.,
  {Henson} C.~N., {Thompson} K.~L., 2004, \apj, 617, 829

\bibitem[{{Benson} {et~al}\mbox{.}(2003){Benson}, {Church}, {Ade}, {Bock},
  {Ganga}, {Hinderks}, {Mauskopf}, {Philhour}, {Runyan}, \&
  {Thompson}}]{Benson2003}
{Benson} B.~A. {et~al.}, 2003, \apj, 592, 674

\bibitem[{{Birkinshaw}(1999)}]{Birkinshaw1999}
{Birkinshaw} M., 1999, \physrep, 310, 97

\bibitem[{{Birkinshaw} {et~al}\mbox{.}(1984){Birkinshaw}, {Gull}, \&
  {Hardebeck}}]{Birkinshaw1984}
{Birkinshaw} M., {Gull} S.~F., {Hardebeck} H., 1984, \nat, 309, 34

\bibitem[{{Birkinshaw} {et~al}\mbox{.}(1991){Birkinshaw}, {Hughes}, \&
  {Arnaud}}]{Birkinshaw1991}
{Birkinshaw} M., {Hughes} J.~P., {Arnaud} K.~A., 1991, \apj, 379, 466

\bibitem[{{Carlstrom} {et~al}\mbox{.}(2002){Carlstrom}, {Holder}, \&
  {Reese}}]{Carlstrom2002}
{Carlstrom} J.~E., {Holder} G.~P., {Reese} E.~D., 2002, \araa, 40, 643

\bibitem[{{Cavaliere} \& {Fusco-Femiano}(1978)}]{Cavaliere1978}
{Cavaliere} A., {Fusco-Femiano} R., 1978, \aap, 70, 677

\bibitem[{{Challinor} \& {Lasenby}(1998)}]{Challinor1998}
{Challinor} A., {Lasenby} A., 1998, \apj, 499, 1

\bibitem[{{Challinor} \& {Lasenby}(1999)}]{Challinor1999}
{Challinor} A., {Lasenby} A., 1999, \apj, 510, 930

\bibitem[{{Chluba}(2011)}]{Chluba2011ab}
{Chluba} J., 2011, \mnras, 415, 3227

\bibitem[{{Chluba} {et~al}\mbox{.}(2005){Chluba}, {H{\"u}tsi}, \&
  {Sunyaev}}]{Chluba2005b}
{Chluba} J., {H{\"u}tsi} G., {Sunyaev} R.~A., 2005, \aap, 434, 811

\bibitem[{{Chluba} \& {Mannheim}(2002)}]{Chluba2002}
{Chluba} J., {Mannheim} K., 2002, \aap, 396, 419

\bibitem[{{Chluba} {et~al}\mbox{.}(2012){Chluba}, {Nagai}, {Sazonov}, \&
  {Nelson}}]{ChlubaSZpack}
{Chluba} J., {Nagai} D., {Sazonov} S., {Nelson} K., 2012, \mnras, 426, 510

\bibitem[{{Colafrancesco} {et~al}\mbox{.}(2011){Colafrancesco}, {Marchegiani},
  \& {Buonanno}}]{Colafrancesco2011}
{Colafrancesco} S., {Marchegiani} P., {Buonanno} R., 2011, \aap, 527, L1

\bibitem[{{Colafrancesco} {et~al}\mbox{.}(2003){Colafrancesco}, {Marchegiani},
  \& {Palladino}}]{Colafrancesco2003}
{Colafrancesco} S., {Marchegiani} P., {Palladino} E., 2003, \aap, 397, 27

\bibitem[{{Diego} {et~al}\mbox{.}(2003){Diego}, {Mazzotta}, \&
  {Silk}}]{Diego2003}
{Diego} J.~M., {Mazzotta} P., {Silk} J., 2003, \apjl, 597, L1

\bibitem[{{Dolgov} {et~al}\mbox{.}(2001){Dolgov}, {Hansen}, {Pastor}, \&
  {Semikoz}}]{Dolgov2001}
{Dolgov} A.~D., {Hansen} S.~H., {Pastor} S., {Semikoz} D.~V., 2001, \apj, 554,
  74

\bibitem[{{En{\ss}lin} \& {Kaiser}(2000)}]{Ensslin2000}
{En{\ss}lin} T.~A., {Kaiser} C.~R., 2000, \aap, 360, 417

\bibitem[{{Finoguenov} {et~al}\mbox{.}(2001){Finoguenov}, {Reiprich}, \&
  {B{\"o}hringer}}]{Finoguenov2001}
{Finoguenov} A., {Reiprich} T.~H., {B{\"o}hringer} H., 2001, \aap, 368, 749

\bibitem[{{Fixsen} {et~al}\mbox{.}(1996){Fixsen}, {Cheng}, {Gales}, {Mather},
  {Shafer}, \& {Wright}}]{Fixsen1996}
{Fixsen} D.~J., {Cheng} E.~S., {Gales} J.~M., {Mather} J.~C., {Shafer} R.~A.,
  {Wright} E.~L., 1996, \apj, 473, 576

\bibitem[{{Fixsen} \& {Mather}(2002)}]{Fixsen2002}
{Fixsen} D.~J., {Mather} J.~C., 2002, \apj, 581, 817

\bibitem[{{Foreman-Mackey} {et~al}\mbox{.}(2012){Foreman-Mackey}, {Hogg},
  {Lang}, \& {Goodman}}]{Foreman2012}
{Foreman-Mackey} D., {Hogg} D.~W., {Lang} D., {Goodman} J., 2012,
  ArXiv:1202.3665

\bibitem[{{Gomez} {et~al}\mbox{.}(2004){Gomez}, {Romer}, {Peterson}, {Chase},
  {Runyan}, {Holzapfel}, {Kuo}, {Newcomb}, {Ruhl}, {Goldstein}, \&
  {Lange}}]{Gomez2004}
{Gomez} P. {et~al.}, 2004, in American Institute of Physics Conference Series,
  Vol. 703, Plasmas in the Laboratory and in the Universe: New Insights and New
  Challenges, {Bertin} G., {Farina} D., {Pozzoli} R., eds., pp. 361--366

\bibitem[{{Hand} {et~al}\mbox{.}(2012){Hand}, {Addison}, {Aubourg},
  {Battaglia}, {Battistelli}, {Bizyaev}, {Bond}, {Brewington}, {Brinkmann},
  {Brown}, {Das}, {Dawson}, {Devlin}, {Dunkley}, {Dunner}, {Eisenstein},
  {Fowler}, {Gralla}, {Hajian}, {Halpern}, {Hilton}, {Hincks}, {Hlozek},
  {Hughes}, {Infante}, {Irwin}, {Kosowsky}, {Lin}, {Malanushenko},
  {Malanushenko}, {Marriage}, {Marsden}, {Menanteau}, {Moodley}, {Niemack},
  {Nolta}, {Oravetz}, {Page}, {Palanque-Delabrouille}, {Pan}, {Reese},
  {Schlegel}, {Schneider}, {Sehgal}, {Shelden}, {Sievers}, {Sif{\'o}n},
  {Simmons}, {Snedden}, {Spergel}, {Staggs}, {Swetz}, {Switzer}, {Trac},
  {Weaver}, {Wollack}, {Yeche}, \& {Zunckel}}]{Hand2012}
{Hand} N. {et~al.}, 2012, Physical Review Letters, 109, 041101

\bibitem[{{Hansen} {et~al}\mbox{.}(2002){Hansen}, {Pastor}, \&
  {Semikoz}}]{Hansen2002}
{Hansen} S.~H., {Pastor} S., {Semikoz} D.~V., 2002, \apjl, 573, L69

\bibitem[{{Hughes} \& {Birkinshaw}(1998)}]{Hughes1998}
{Hughes} J.~P., {Birkinshaw} M., 1998, \apj, 501, 1

\bibitem[{{Itoh} {et~al}\mbox{.}(2001){Itoh}, {Kawana}, {Nozawa}, \&
  {Kohyama}}]{Itoh2001}
{Itoh} N., {Kawana} Y., {Nozawa} S., {Kohyama} Y., 2001, \mnras, 327, 567

\bibitem[{{Itoh} {et~al}\mbox{.}(1998){Itoh}, {Kohyama}, \& {Nozawa}}]{Itoh98}
{Itoh} N., {Kohyama} Y., {Nozawa} S., 1998, \apj, 502, 7

\bibitem[{{Itoh} \& {Nozawa}(2004)}]{Itoh2004fittingII}
{Itoh} N., {Nozawa} S., 2004, \aap, 417, 827

\bibitem[{{Kitayama} {et~al}\mbox{.}(2004){Kitayama}, {Komatsu}, {Ota},
  {Kuwabara}, {Suto}, {Yoshikawa}, {Hattori}, \& {Matsuo}}]{Kitayama2004}
{Kitayama} T., {Komatsu} E., {Ota} N., {Kuwabara} T., {Suto} Y., {Yoshikawa}
  K., {Hattori} M., {Matsuo} H., 2004, \pasj, 56, 17

\bibitem[{{Komatsu} {et~al}\mbox{.}(2001){Komatsu}, {Matsuo}, {Kitayama},
  {Hattori}, {Kawabe}, {Kohno}, {Kuno}, {Schindler}, {Suto}, \&
  {Yoshikawa}}]{Komatsu2001}
{Komatsu} E. {et~al.}, 2001, \pasj, 53, 57

\bibitem[{{Komatsu} \& {Seljak}(2002)}]{Komatsu2002}
{Komatsu} E., {Seljak} U., 2002, \mnras, 336, 1256

\bibitem[{{Korngut} {et~al}\mbox{.}(2011){Korngut}, {Dicker}, {Reese}, {Mason},
  {Devlin}, {Mroczkowski}, {Sarazin}, {Sun}, \& {Sievers}}]{Korngut2011}
{Korngut} P.~M. {et~al.}, 2011, \apj, 734, 10

\bibitem[{Kravtsov {et~al}\mbox{.}(2002)Kravtsov, Klypin, \&
  Hoffman}]{Kravtsov2002}
Kravtsov A.~V., Klypin A., Hoffman Y., 2002, \apj, 571, 563

\bibitem[{{Lamarre} {et~al}\mbox{.}(1998){Lamarre}, {Giard}, {Pointecouteau},
  {Bernard}, {Serra}, {Pajot}, {D{\'e}sert}, {Ristorcelli}, {Torre}, {Church},
  {Coron}, {Puget}, \& {Bock}}]{Lamarre1998}
{Lamarre} J.~M. {et~al.}, 1998, \apjl, 507, L5

\bibitem[{{Markevitch} \& {Vikhlinin}(2007)}]{Markevitch2007}
{Markevitch} M., {Vikhlinin} A., 2007, \physrep, 443, 1

\bibitem[{{Markevitch} {et~al}\mbox{.}(1999){Markevitch}, {Vikhlinin},
  {Forman}, \& {Sarazin}}]{Markevitch1999}
{Markevitch} M., {Vikhlinin} A., {Forman} W.~R., {Sarazin} C.~L., 1999, \apj,
  527, 545

\bibitem[{{Marriage} {et~al}\mbox{.}(2011){Marriage}, {Acquaviva}, {Ade},
  {Aguirre}, {Amiri}, {Appel}, {Barrientos}, {Battistelli}, {Bond}, {Brown},
  {Burger}, {Chervenak}, {Das}, {Devlin}, {Dicker}, {Bertrand Doriese},
  {Dunkley}, {D{\"u}nner}, {Essinger-Hileman}, {Fisher}, {Fowler}, {Hajian},
  {Halpern}, {Hasselfield}, {Hern{\'a}ndez-Monteagudo}, {Hilton}, {Hilton},
  {Hincks}, {Hlozek}, {Huffenberger}, {Handel Hughes}, {Hughes}, {Infante},
  {Irwin}, {Baptiste Juin}, {Kaul}, {Klein}, {Kosowsky}, {Lau}, {Limon}, {Lin},
  {Lupton}, {Marsden}, {Martocci}, {Mauskopf}, {Menanteau}, {Moodley},
  {Moseley}, {Netterfield}, {Niemack}, {Nolta}, {Page}, {Parker}, {Partridge},
  {Quintana}, {Reese}, {Reid}, {Sehgal}, {Sherwin}, {Sievers}, {Spergel},
  {Staggs}, {Swetz}, {Switzer}, {Thornton}, {Trac}, {Tucker}, {Warne},
  {Wilson}, {Wollack}, \& {Zhao}}]{Marriage2011}
{Marriage} T.~A. {et~al.}, 2011, \apj, 737, 61

\bibitem[{{Menanteau} {et~al}\mbox{.}(2012){Menanteau}, {Hughes}, {Sif{\'o}n},
  {Hilton}, {Gonz{\'a}lez}, {Infante}, {Barrientos}, {Baker}, {Bond}, {Das},
  {Devlin}, {Dunkley}, {Hajian}, {Hincks}, {Kosowsky}, {Marsden}, {Marriage},
  {Moodley}, {Niemack}, {Nolta}, {Page}, {Reese}, {Sehgal}, {Sievers},
  {Spergel}, {Staggs}, \& {Wollack}}]{ElGordo2012}
{Menanteau} F. {et~al.}, 2012, \apj, 748, 7

\bibitem[{{Million} \& {Allen}(2009)}]{Million2009}
{Million} E.~T., {Allen} S.~W., 2009, \mnras, 399, 1307

\bibitem[{{Mroczkowski} {et~al}\mbox{.}(2012){Mroczkowski}, {Dicker}, {Sayers},
  {Reese}, {Mason}, {Czakon}, {Romero}, {Young}, {Devlin}, {Golwala},
  {Korngut}, {Sarazin}, {Bock}, {Koch}, {Lin}, {Molnar}, {Pierpaoli}, {Umetsu},
  \& {Zemcov}}]{Mroczkowski2012}
{Mroczkowski} T. {et~al.}, 2012, \apj, 761, 47

\bibitem[{{Nagai} {et~al}\mbox{.}(2003){Nagai}, {Kravtsov}, \&
  {Kosowsky}}]{Nagai2003}
{Nagai} D., {Kravtsov} A.~V., {Kosowsky} A., 2003, \apj, 587, 524

\bibitem[{Nagai {et~al}\mbox{.}(2007{\natexlab{a}})Nagai, Kravtsov, \&
  Vikhlinin}]{Nagai2007b}
Nagai D., Kravtsov A.~V., Vikhlinin A., 2007{\natexlab{a}}, \apj, 668, 1

\bibitem[{Nagai {et~al}\mbox{.}(2007{\natexlab{b}})Nagai, Vikhlinin, \&
  Kravtsov}]{Nagai2007a}
Nagai D., Vikhlinin A., Kravtsov A.~V., 2007{\natexlab{b}}, \apj, 655, 98

\bibitem[{{Nelson} {et~al}\mbox{.}(2012){Nelson}, {Rudd}, {Shaw}, \&
  {Nagai}}]{Nelson2011}
{Nelson} K., {Rudd} D.~H., {Shaw} L., {Nagai} D., 2012, \apj, 751, 121

\bibitem[{{Nozawa} {et~al}\mbox{.}(2000){Nozawa}, {Itoh}, {Kawana}, \&
  {Kohyama}}]{Nozawa2000fitting}
{Nozawa} S., {Itoh} N., {Kawana} Y., {Kohyama} Y., 2000, \apj, 536, 31

\bibitem[{{Nozawa} {et~al}\mbox{.}(1998{\natexlab{a}}){Nozawa}, {Itoh}, \&
  {Kohyama}}]{Nozawa1998SZ}
{Nozawa} S., {Itoh} N., {Kohyama} Y., 1998{\natexlab{a}}, \apj, 508, 17

\bibitem[{{Nozawa} {et~al}\mbox{.}(1998{\natexlab{b}}){Nozawa}, {Itoh}, \&
  {Kohyama}}]{Nozawa1998}
{Nozawa} S., {Itoh} N., {Kohyama} Y., 1998{\natexlab{b}}, \apj, 507, 530

\bibitem[{{Nozawa} {et~al}\mbox{.}(2009){Nozawa}, {Kohyama}, \&
  {Itoh}}]{Nozawa2009}
{Nozawa} S., {Kohyama} Y., {Itoh} N., 2009, \prd, 79, 123007

\bibitem[{{Planck Collaboration} {et~al}\mbox{.}(2011){Planck Collaboration},
  {Ade}, {Aghanim}, {Arnaud}, {Ashdown}, {Aumont}, {Baccigalupi}, {Balbi},
  {Banday}, {Barreiro}, \& et~al.}]{AdeESZCS}
{Planck Collaboration} {et~al.}, 2011, \aap, 536, A8

\bibitem[{{Pointecouteau} {et~al}\mbox{.}(1998){Pointecouteau}, {Giard}, \&
  {Barret}}]{Pointecouteau1998}
{Pointecouteau} E., {Giard} M., {Barret} D., 1998, \aap, 336, 44

\bibitem[{{Poutanen} \& {Vurm}(2010)}]{Poutanen2010}
{Poutanen} J., {Vurm} I., 2010, \apjs, 189, 286

\bibitem[{{Pratt} \& {Arnaud}(2002)}]{Pratt2002}
{Pratt} G.~W., {Arnaud} M., 2002, \aap, 394, 375

\bibitem[{{Prokhorov} \& {Colafrancesco}(2012)}]{Prokhorov2012}
{Prokhorov} D.~A., {Colafrancesco} S., 2012, \mnras, 424, L49

\bibitem[{{Prokhorov} {et~al}\mbox{.}(2011){Prokhorov}, {Colafrancesco},
  {Akahori}, {Million}, {Nagataki}, \& {Yoshikawa}}]{Prokhorov2011}
{Prokhorov} D.~A., {Colafrancesco} S., {Akahori} T., {Million} E.~T.,
  {Nagataki} S., {Yoshikawa} K., 2011, \mnras, 416, 302

\bibitem[{{Reese} {et~al}\mbox{.}(2002){Reese}, {Carlstrom}, {Joy}, {Mohr},
  {Grego}, \& {Holzapfel}}]{Reese2002}
{Reese} E.~D., {Carlstrom} J.~E., {Joy} M., {Mohr} J.~J., {Grego} L.,
  {Holzapfel} W.~L., 2002, \apj, 581, 53

\bibitem[{{Rephaeli}(1995{\natexlab{a}})}]{Rephaeli1995ARAA}
{Rephaeli} Y., 1995{\natexlab{a}}, \araa, 33, 541

\bibitem[{{Rephaeli}(1995{\natexlab{b}})}]{Rephaeli1995}
{Rephaeli} Y., 1995{\natexlab{b}}, \apj, 445, 33

\bibitem[{{Rudd} \& {Nagai}(2009)}]{Rudd2009}
{Rudd} D.~H., {Nagai} D., 2009, \apjl, 701, L16

\bibitem[{Rudd {et~al}\mbox{.}(2008)Rudd, Zentner, \& Kravtsov}]{Rudd2008}
Rudd D.~H., Zentner A.~R., Kravtsov A.~V., 2008, \apj, 672, 19

\bibitem[{{Sazonov} \& {Sunyaev}(1998)}]{Sazonov1998}
{Sazonov} S.~Y., {Sunyaev} R.~A., 1998, \apj, 508, 1

\bibitem[{{Shimon} \& {Rephaeli}(2004)}]{Shimon2004}
{Shimon} M., {Rephaeli} Y., 2004, New Astronomy, 9, 69

\bibitem[{{Sunyaev} {et~al}\mbox{.}(2003){Sunyaev}, {Norman}, \&
  {Bryan}}]{Sunyaev2003}
{Sunyaev} R.~A., {Norman} M.~L., {Bryan} G.~L., 2003, Astronomy Letters, 29,
  783

\bibitem[{{Sunyaev} \& {Zeldovich}(1980)}]{Sunyaev1980}
{Sunyaev} R.~A., {Zeldovich} I.~B., 1980, \mnras, 190, 413

\bibitem[{{Vikhlinin} {et~al}\mbox{.}(2006){Vikhlinin}, {Kravtsov}, {Forman},
  {Jones}, {Markevitch}, {Murray}, \& {Van Speybroeck}}]{Vikhlinin2006}
{Vikhlinin} A., {Kravtsov} A., {Forman} W., {Jones} C., {Markevitch} M.,
  {Murray} S.~S., {Van Speybroeck} L., 2006, \apj, 640, 691

\bibitem[{{Williamson} {et~al}\mbox{.}(2011){Williamson}, {Benson}, {High},
  {Vanderlinde}, {Ade}, {Aird}, {Andersson}, {Armstrong}, {Ashby}, {Bautz},
  {Bazin}, {Bertin}, {Bleem}, {Bonamente}, {Brodwin}, {Carlstrom}, {Chang},
  {Chapman}, {Clocchiatti}, {Crawford}, {Crites}, {de Haan}, {Desai}, {Dobbs},
  {Dudley}, {Fazio}, {Foley}, {Forman}, {Garmire}, {George}, {Gladders},
  {Gonzalez}, {Halverson}, {Holder}, {Holzapfel}, {Hoover}, {Hrubes}, {Jones},
  {Joy}, {Keisler}, {Knox}, {Lee}, {Leitch}, {Lueker}, {Luong-Van}, {Marrone},
  {McMahon}, {Mehl}, {Meyer}, {Mohr}, {Montroy}, {Murray}, {Padin}, {Plagge},
  {Pryke}, {Reichardt}, {Rest}, {Ruel}, {Ruhl}, {Saliwanchik}, {Saro},
  {Schaffer}, {Shaw}, {Shirokoff}, {Song}, {Spieler}, {Stalder}, {Stanford},
  {Staniszewski}, {Stark}, {Story}, {Stubbs}, {Vieira}, {Vikhlinin}, \&
  {Zenteno}}]{Williamson2011}
{Williamson} R. {et~al.}, 2011, \apj, 738, 139

\bibitem[{{Wright}(1979)}]{Wright1979}
{Wright} E.~L., 1979, \apj, 232, 348

\bibitem[{{Zeldovich} \& {Sunyaev}(1969)}]{Zeldovich1969}
{Zeldovich} Y.~B., {Sunyaev} R.~A., 1969, \apss, 4, 301

\bibitem[{{Zemcov} {et~al}\mbox{.}(2012){Zemcov}, {Aguirre}, {Bock},
  {Bradford}, {Czakon}, {Glenn}, {Golwala}, {Lupu}, {Maloney}, {Mauskopf},
  {Million}, {Murphy}, {Naylor}, {Nguyen}, {Rosenman}, {Sayers}, {Scott}, \&
  {Zmuidzinas}}]{Zemcov2012}
{Zemcov} M. {et~al.}, 2012, \apj, 749, 114

\bibitem[{{Zemcov} {et~al}\mbox{.}(2010){Zemcov}, {Rex}, {Rawle}, {Bock},
  {Egami}, {Altieri}, {Blain}, {Boone}, {Bridge}, {Clement}, {Combes},
  {Dowell}, {Dessauges-Zavadsky}, {Fadda}, {Ilbert}, {Ivison}, {Jauzac},
  {Kneib}, {Lutz}, {Pell{\'o}}, {Pereira}, {P{\'e}rez-Gonz{\'a}lez}, {Richard},
  {Rieke}, {Rodighiero}, {Schaerer}, {Smith}, {Valtchanov}, {Walth}, {van der
  Werf}, \& {Werner}}]{Zemcov2010}
{Zemcov} M. {et~al.}, 2010, \aap, 518, L16

\end{thebibliography}
